\documentclass[12pt]{iopart}
\usepackage{iopams}
\usepackage{graphicx}
\usepackage{color}
\usepackage{epsfig}
\def\la{\left\langle}
\def\ra{\right\rangle}
\begin{document}
\title{The Universality of Dynamic Multiscaling in Homogeneous, 
Isotropic Turbulence}
\author{Samriddhi Sankar Ray$^{1}$, 
Dhrubaditya Mitra$^2$\footnote{Now at: Astronomy Unit, School of 
Mathematical Sciences, Queen Mary College, London, United Kingdom. }
and Rahul Pandit$^{1}$\footnote{Author to whom all 
correspondence should be addressed.}\footnote{Also at: Jawaharlal Nehru Centre for 
Advanced Scientific Research, Bangalore, India.}}
\address{$^1$ Centre for Condensed Matter Theory, Department of
Physics, Indian Institute of Science, Bangalore, India.\\
$^2$D\'epartement Cassiop\'ee, Observatoire de la C\^{o}te d'Azur,
BP4229, 06304 Nice Cedex 4, France.\\} 
\ead{rahul@physics.iisc.ernet.in} 
\begin{abstract}
We systematise the study of dynamic multiscaling 
of time-dependent structure functions in different models 
of passive-scalar and fluid turbulence.  
We show that, by suitably normalising these structure functions, 
we can eliminate their dependence on the origin of time at which
we start our measurements and that these normalised structure 
functions yield the same linear bridge relations that relate the 
dynamic-multiscaling and equal-time exponents for statistically 
steady turbulence. We show analytically, for both the Kraichnan 
Model of passive-scalar turbulence and its shell model analogue, 
and numerically, for the GOY shell model of fluid turbulence and 
a shell model for passive-scalar turbulence, that these exponents 
and bridge relations are the same for statistically steady and 
decaying turbulence. Thus we provide strong evidence for dynamic 
universality, i.e., dynamic-multiscaling exponents do not depend 
on whether the turbulence decays or is statistically steady.
\end{abstract}
\noindent{\it Turbulence, Multifractality, Dynamic Scaling}
\maketitle
\section{Introduction}
\label{intro}
The elucidation of the universal scaling properties of 
equal-time and time-dependent correlation functions in the 
vicinity of a critical point was one of the most important 
achievements of statistical mechanics over the past forty 
years. The analogous systematization of the power 
laws and associated exponents that govern the behaviours of 
structure functions in a turbulent fluid, or in a passive-scalar 
advected by such a fluid, is a major challenge
in the areas of nonequilibrium statistical mechanics, fluid 
mechanics, and nonlinear dynamics. The power-law behaviours of
{\it equal-time} structure functions have been studied in detail 
over the past few decades \cite{frisch}; and, especially in the 
case of passive-scalar turbulence \cite{falcormp}, significant 
progress has been made in understanding the {\it multiscaling} of 
{\it equal-time} structure functions. The nature of multiscaling of 
{\it time-dependent} structure functions has been examined 
recently \cite{lvov,bif,mitra0,mitra1,mitra2} but only for the case 
of {\it statistically steady} turbulence. We develop here the 
systematics of the multiscaling of time-dependent structure 
functions for the case of {\it decaying} fluid and passive-scalar 
turbulence \cite{foot1}.

To set the stage for our discussion of time-dependent structure 
functions in turbulence, it is useful to begin by recalling some 
well-known results from critical phenomena~\cite{rmp,chaikin}: At a 
critical point for a spin system in $d$ dimensions, the equal-time, 
two-spin correlation function $g$ and its spatial Fourier 
transform $\tilde g$ assume the following power-law scaling 
forms: 
\begin{eqnarray}
g({\bf r}; \bar{t},h) &\approx& 
\frac{G(r\bar{t}^{\nu},h/\bar{t}^{\Delta})}{r^{d-2+\eta}}\/; 
\nonumber \\
\tilde g({\bf k}; \bar{t},h)  &\approx& 
\frac{{\tilde G}(k/\bar{t}^{\nu},h/\bar{t}^{\Delta})}{k^{2-\eta}}
 \/. 
\end{eqnarray}
Here $\bar{t}\equiv (|T - T_c|)/T_c$, $T$ and $T_c$ are the 
temperature and the critical temperature, respectively, 
$h \equiv H/k_BT_c$, $H$ is the external field, $k_B$ is the 
Boltzmann constant, the spins are separated by the vector 
${\bf r}$ [$r=|{\bf r}|$], $\bf{k}$ is the wavevector, 
$k=|{\bf k}|$, $\nu$, $\Delta$, $\eta$ are critical exponents, and 
$G$ and ${\tilde G}$ are scaling functions. Away from the 
critical point such correlation functions decay exponentially; the 
associated correlation length $\xi_c$ diverges in the vicinity of the 
critical point; e.g., as $\xi_c \sim \bar{t}^{-\nu}$, if $h = 0$. 
Time-dependent correlation functions also assume scaling forms in 
the vicinity of the critical point and the characteristic 
relaxation time $\tau$ diverges as suggested by the 
{\it dynamic-scaling Ansatz} \cite{rmp}   
\begin{equation}
\tau \sim \xi_c^{z},  
\end{equation}
which introduces the {\it dynamic-scaling exponent} $z$. 

The generalisation of such a dynamic-scaling {\it Ansatz} to the case of 
homogeneous, isotropic turbulence is our prime concern here. The 
power-law behaviours of equal-time structure functions, in the inertial 
range (to be defined later), in turbulence are reminiscent of the 
algebraic dependence on $r$ of critical-point correlation functions. 
However, there are important differences between the two that must be 
appreciated before we embark on a systematization of time-dependent 
structure functions in turbulence.  We begin with the increments of  
the longitudinal component of the velocity  
$\delta u_{\parallel}({\bf x},{\bf r},t) \equiv [{\bf u}({\bf x}+
{\bf r},t) - {\bf u}({\bf x},t)]\cdot({\bf r}/r)$ and passive-scalar 
$\delta\theta({\bf x},{\bf r},t) \equiv [\theta({\bf x} + 
{\bf r},t) - \theta({\bf x},t)]$, respectively; here 
${\bf u}({\bf x},t)$ and $\theta({\bf x},t)$ denote, respectively, 
the velocity of the fluid and the passive-scalar density at the 
point ${\bf x}$ and time $t$, and the subscript $\parallel$ the 
longitudinal component. The order-$p$, equal-time structure 
functions, for the fluid (superscript $u$) and passive-scalar 
(superscript $\theta$) fields, are defined as follows: 
\begin{eqnarray}
{\cal S}^{u}_p(r) &\equiv& \la [\delta u_{\parallel}
({\bf x},{\bf r},t)]^p \ra 
\sim r^{\zeta^{u}_p} ; \nonumber \\
{\cal S}^{\theta}_p(r) &\equiv&
                \la [\delta\theta({\bf x},{\bf r},t)]^p \ra 
\sim r^{\zeta^{\theta}_p};
\label{Sp}
\end{eqnarray}
the angular brackets indicate averages over the steady state for 
statistically steady turbulence or over statistically independent 
initial configurations for decaying turbulence; for stochastic 
differential equations, like the Kraichnan Model ($\Sref{kraichnan}$),
the angular brackets denote an average over the statistics of the 
noise; and the power laws, characterised by the equal-time 
exponents $\zeta^{u}_p$ and $\zeta^{\theta}_p$, hold for 
separations $r$ in the inertial range $\eta_d \ll r \ll L$, where 
$\eta_d$ is the Kolmogorov dissipation scale and $L$ the large 
length scale at which energy is injected into the system.  

Kolmogorov's phenomenological theory \cite{frisch,K41a, K41b} of 
1941 (K41) suggests simple scaling, with  $\zeta^{u,K41}_p = p/3$, 
but experimental and numerical evidence favours equal-time 
multiscaling with $\zeta^{u}_p$ and $\zeta^{\theta}_p$ nonlinear, 
convex, monotone-increasing functions of $p$. For the simplified 
Kraichnan model \cite{falcormp,kraich1,kraich2,kraich3} 
of passive-scalar turbulence
($\Sref{kraichnan}$) multiscaling of equal-time structure functions 
can be demonstrated analytically in certain limits. The analogue
of the K41 theory for passive-scalar turbulence is due to Obukhov
and Corrsin \cite{obu,corr}; if the Schmidt number 
$Sc \equiv \nu/\kappa \simeq 1$, then their theory yields K41 
exponents for the passive-scalar case; here $\nu$ is the kinematic 
viscosity of the fluid and $\kappa$ is the diffusivity of the 
passive scalar. 

A straightforward extension of simple, K41 scaling to 
time-dependent structure functions implies that the dynamic 
exponents $z^{K41}_p = 2/3$ for all $p$. This na\"ive extension 
fails for two reasons: (a) it does not distinguish between the 
temporal behaviours of structure functions of Eulerian, 
Lagrangian, and quasi-Lagrangian ($\Sref{models}$) velocities or 
passive-scalar densities; and (b) it does not account for the 
multiscaling of structure functions. These difficulties have been 
overcome to a large extent for statistically steady 
turbulence \cite{lvov,bif,mitra0,mitra1,mitra2,kaneda} as we 
summarise below. There is consensus now that Eulerian structure 
functions display simple scaling with only one dynamic-scaling 
exponent $z^{\cal{E}}=1$ because of the {\it sweeping effect}: the 
mean flow, or the flow caused by the largest eddy, advects small 
eddies, so spatial separations $r$ in (\ref{Sp}) are related 
{\it linearly} to temporal separations $\tau$ via the mean-flow 
velocity \cite{kaneda}. By contrast, it is expected that 
Lagrangian \cite{kaneda} or quasi-Lagrangian \cite{lvov,bif,mitra0,
mitra1,mitra2,belinicher} 
time-dependent structure functions should show nontrivial dynamic 
multiscaling. The task of extracting well-averaged time-dependent
Lagrangian or quasi-Lagrangian structure functions from a 
direct numerical simulation (DNS) of the Navier-Stokes 
equation is a daunting one \cite{mitra}: a dynamic exponent has been 
extracted from a full Lagrangian study \cite{kaneda} only for 
order $p = 2$.
Thus the elucidation of dynamic multiscaling has relied on 
predictions based on generalisations of the multifractal 
formalism \cite{lvov,bif,mitra0,mitra1,mitra2} and on numerical 
studies of shell models \cite{bif,mitra0,mitra1,mitra2}.
These studies show that, if dynamic multiscaling exists,
time-dependent structure functions must be characterised by an 
infinity of time scales and associated dynamic multiscaling 
exponents \cite{mitra1}. Furthermore, the dynamic exponents depend 
on how we extract time scales from time-dependent structure 
functions; 
e.g., for fluid turbulence, time scales obtained from integrals 
(superscript $I$ and subscript 1) and second derivatives 
(superscript $D$ and subscript $2$) of order-$p$ time-dependent 
structure functions yield the {\it different} dynamic exponents 
$z_{p,1}^{I,u}$ and $z_{p,2}^{D,u}$. 
Finally, the different dynamic multiscaling exponents are related 
by   {\it different classes of linear bridge relations} to the 
equal-time multiscaling exponents. For a careful discussion of these
issues we must of course define time-dependent structure functions.
The details necessary for this paper are given in $\Sref{models}$ and 
$\Sref{multifractal}$.

The dynamic multiscaling of time-dependent structure 
functions described briefly above applies to statistically 
steady turbulence. Does it have an analogue in the case of 
decaying turbulence, since time-dependent structure functions 
must, in this case, depend on the origin of time $t_0$ at which 
we start our measurements? This question has not been addressed 
hitherto. We show here how to answer it in decaying fluid and 
passive-scalar turbulence \cite{foot1}. In particular, we propose 
suitable normalisations of time-dependent structure functions that 
eliminate their dependence on $t_0$; we demonstrate this 
{\it analytically} for the Kraichnan version of the passive-scalar 
problem and its shell-model analogue and {\it numerically} for the 
GOY shell model \cite{frisch,gledzer,ohkitani} for
fluids and a shell-model version of the advection-diffusion 
equation. In these models we then analyse the normalised 
time-dependent structure functions for the case of decaying 
turbulence like their statistically steady 
counterparts \cite{mitra1,mitra2}. This requires a generalisation of 
the multifractal formalism \cite{frisch} that finally yields the 
same bridge relations between dynamic and equal-time multiscaling 
exponents as  for statistically steady 
turbulence \cite{mitra1,mitra2}. For the Kraichnan version of the 
passive-scalar problem we show analytically that simple dynamic 
scaling is obtained. This is because ($\Sref{analytic}$) the advecting velocity 
is random and white in time. In addition, we find numerically for 
shell models of fluid and passive-scalar turbulence that 
dynamic-multiscaling exponents have the same values for both 
statistically steady and decaying turbulence; so, 
in this sense, we have {\it universality} of the multiscaling of 
time-dependent structure functions in turbulence. The equal-time 
analogue of this universality has been discussed in 
Ref.\cite{lvov1}.

The remaining part of this paper is organized as follows. In 
$\Sref{models}$ we introduce the models we use and give the details 
of our numerical simulations. $\Sref{analytic}$ presents our 
analytical studies of decaying turbulence in the Kraichnan model 
and its shell-model analogue. $\Sref{multifractal}$ shows how to 
generalise the multifractal formalism to allow for time-dependent
structure functions in decaying turbulence and how to obtain
bridge relations between dynamic and equal-time multiscaling 
exponents in this case. In $\Sref{results}$ we present the results 
of our numerical studies of dynamic multiscaling in the GOY shell 
model 
for fluid turbulence and for shell models of a passive-scalar field 
advected by a turbulent velocity field. $\Sref{conclusion}$ is 
devoted to a 
discussion of our results in the context of earlier studies; we 
also suggest possible experimental tests of our predictions. 
 
\section{Models and Numerical Simulations}
\label{models}
	We have used several models to study time-dependent 
structure functions in fluid and passive-scalar turbulence. These 
range from the Navier-Stokes and advection-diffusion equations to 
simple shell models; the latter are well-suited for our extensive 
numerical studies.  It is useful to begin with a systematic 
description of these models.

Fluid flows are governed by the Navier-Stokes (NS) equation 
(\ref{NS1}) for the velocity field ${\bf u}({\bf x},t)$ at point 
${\bf x}$ and time $t$, augmented by the incompressibility 
constraint (\ref{NS2}), since we restrict ourselves to low Mach 
numbers: 
\begin{equation}
\partial_t{\bf u} + {\bf u.}\nabla{\bf u} = -\nabla P + 
\nu_0\nabla^2{\bf u} + {\bf f} ; \\
\label{NS1}
\end{equation}
\begin{equation}
\nabla{\bf .u} = 0.
\label{NS2}
\end{equation}
Here $\nu_0$ is the kinematic viscosity, $P$ the pressure, the density 
$\rho$ is taken to be 1, and ${\bf f}$ the external force, which is 
absent when we consider decaying turbulence.  If $\ell$ and $v$ are, 
respectively, characteristic length and velocity scales of the flow, 
the Reynolds number $Re \equiv \frac{\ell v}{\nu_0}$ provides a 
dimensionless measure of the strength of the nonlinear term in 
(\ref{NS1}) relative to the viscous term; for the case of decaying 
turbulence it is convenient to use the Reynolds number for the 
initial state, i.e., $Re$ with $v$ the root-mean-square (rms) 
velocity of the initial condition and $\ell$ the system size [the 
linear size of the simulation box in a direct numerical simulation 
(DNS)]. Given the incompressibility condition (\ref{NS2}), the 
pressure can be eliminated from (\ref{NS1}) and related to the 
velocity by a Poisson equation. The equation for the velocity alone 
is most easily written in terms of the spatial Fourier transform 
$\tilde{\bf u}({\bf k},t)$ 
of ${\bf u}({\bf x},t)$; and it can be shown easily 
that $\tilde{\bf u}({\bf k},t)$ is affected {\it directly} by all 
other Fourier modes. This is the mathematical representation of the 
sweeping effect in which the largest eddies (i.e., modes with small 
$k \equiv \mid{\bf k}\mid$) {\it directly} advect the 
smallest eddies (i.e., large-$k$ modes); such direct sweeping lies 
at the heart of the Taylor hypothesis \cite{taylor} and leads 
eventually to trivial dynamic scaling for time-dependent structure 
functions of Eulerian fields with a dynamic exponent 
$z^{\cal E} = 1$ for fluid turbulence.

As we have mentioned above, nontrivial dynamic multiscaling is 
expected if we use Lagrangian or quasi-Lagrangian velocities. The 
Lagrangian formulation is well known \cite{pope}; the 
quasi-Lagrangian \cite{lvov,belinicher} one uses the following 
transformation for any Eulerian field $\psi({\bf x},t)$:   
\begin{equation}
{\hat \psi}({\bf x},t) 
\equiv \psi[{\bf x} + {\bf R}(t;{\bf r_0},0),t] ,
\label{qltrans}
\end{equation}
where ${\hat \psi}$ is the quasi-Lagrangian field and
${\bf R}(t;{\bf r_0},0)$ is the position at time $t$ of a 
Lagrangian particle that was at point ${\bf r_0}$ at time $t = 0$.

The advection-diffusion (AD) equation for the Eulerian 
passive-scalar field $\theta({\bf x},t)$ 
is
\begin{equation}
\frac{\partial\theta}{\partial t} + {\bf u}.\nabla \theta = 
\kappa\nabla^2\theta + f_\theta,
\label{nseqn}
\end{equation}
where $\kappa$ is the passive-scalar diffusivity and,  
if we consider decaying passive-scalar turbulence, 
the external force $f_\theta$ is set to zero.  
The advecting velocity field  ${\bf u}$ should be obtained,
in principle, by solving equations (\ref{NS1}) and (\ref{NS2}).  
By using equation (\ref{qltrans}) we get the quasi-Lagrangian 
version of the advection-diffusion equation (\ref{nseqn}):
\begin{equation}
\frac{\partial{\hat \theta}({\bf x},t)}{\partial t} +
\left[{\bf {\hat u}}({\bf x},t) - 
{\bf{\hat  u}}({\bf x},0)\right].\nabla {\hat \theta}({\bf x},t)
  = \kappa \nabla^2 {\hat \theta}({\bf x},t) + 
{\hat f}_{\theta}({\bf x},t).
\label{eq:pas:qlpasscal}
\end{equation}

Direct numerical simulations of equations (\ref{NS1}) and (\ref{NS2})
or equation (\ref{nseqn}), though feasible, have not yet provided
data that are averaged well enough to yield reliable 
{\it time-dependent} structure functions of quasi-Lagrangian 
velocity \cite{mitra} or passive-scalar fields. Time-dependent 
Lagrangian structure functions have been obtained \cite{kaneda} only 
for order $p = 2$. Thus a first-principles DNS study of dynamic 
multiscaling in fluid or passive-scalar turbulence is not possible 
at the moment.  However, significant progress has been made 
in statistically steady turbulence by studying dynamic multiscaling 
in simplified models like the Kraichnan model for 
passive-scalar turbulence and shell models for fluid and 
passive-scalar turbulence. We discuss these models below since our 
studies of decaying turbulence will be based on them.

\subsection{The Kraichnan Model (Model A)}
\label{kraichnan}
The Kraichnan model for passive-scalar turbulence 
 \cite{falcormp,kraich1,kraich2,kraich3} begins 
with the advection-diffusion equation (\ref{nseqn}) but replaces 
the Navier-Stokes velocity field by one in which each 
component $u_i({\bf x},t)$ of the velocity is a zero-mean, 
delta-correlated, Gaussian random variable with the covariance 
\begin{equation}
\la u_i({\bf x},t)u_j({\bf x} + {\bf r},t^{\prime})\ra = 
2D_{ij}({\bf r})\delta(t-t^{\prime}). 
\label{covarnce}
\end{equation}
The Fourier transform of $D_{ij}({\bf r})$ has the form 
\begin{equation}
\tilde D_{ij}({\bf q}) \propto \left(q^2 + 
\frac{1}{L^2}\right)^{-(d + \xi)/2}e^{-\eta q^2}
\left[\delta_{ij} - \frac{q_iq_j}{q^2}\right],
\label{Dij1}
\end{equation}
where {\bf q} is the wave vector, $L$ the characteristic 
large length scale, $\eta_d$ the dissipation scale, and $\xi$ a 
tunable parameter. In the limits $L \rightarrow \infty$ and 
$\eta_d \rightarrow 0$, of relevance to turbulence, we have, in real 
space,
\begin{equation}
D_{ij}({\bf r}) = D^0\delta_{ij} - \frac {1}{2}d_{ij}({\bf r}),
\label{Dij_asymp}
\end{equation}
with
\begin{equation}
d_{ij} = D_1r^\xi\left[(d-1+\xi)\delta_{ij} -
\xi\frac{r_ir_j}{r^2}\right],
\label{dij}
\end{equation}
where $D^0 \sim C_1L^{\xi}$; $C_1$ and $D_1$ are dimensional 
constants.  We refer to (\ref{nseqn}-\ref{dij}) as Model A to 
distinguish it from other models that we use. For $0 < \xi < 2$, 
this model shows multiscaling of order-$p$, equal-time 
passive-scalar structure functions, as can be shown analytically,
in certain limits \cite{falcormp}. However, for the case of 
statistically steady turbulence, this model exhibits simple dynamic 
scaling~\cite{mitra1}.
 
\subsection{Shell Models}
\label{shell}
We will also use some shell models for fluid and passive-scalar
turbulence. These models are highly simplified representations 
of the Navier-Stokes or the advection-diffusion equations 
(\ref{NS1}-\ref{nseqn}) and are, 
therefore, far more tractable numerically than 
(\ref{NS1}-\ref{nseqn}). Nevertheless, shell models retain enough 
properties of their parent equations to make them useful testing 
grounds for the multiscaling of structure functions in turbulence.
Shell models are defined on a logarithmically discretised 
Fourier space in which complex scalar variables (e.g., the velocity 
$u_n$ or passive scalar $\theta_n$) are associated with the shells 
$n$ and scalar wave vectors $k_n = k_0 \lambda^n $; typically 
$\lambda = 2, \; k_0 = 1/16$; and the boundary conditions are 
that the shell variables vanish if $n < 1$ or $n > N$ (we use 
$N = 22$). These models consist of coupled, nonlinear, ordinary 
differential equations (ODEs) that specify the temporal evolution of 
the shell variables $u_n$ and $\theta_n$. Shell-model ODEs are similar 
to the Fourier-space versions of their parent partial differential 
equations:
(a) their dissipative terms are linear in one of the shell variables 
and quadratic in $k_n$; (b) their analogues of advection terms are 
linear in $k_n$ and bilinear in the shell variables; e.g., for fluid 
turbulence a representative term is of the form $ik_n u_nu_{n'}$, 
with $n \neq n'$; and (c) they conserve the shell-model analogues 
of the energy, helicity, etc., in the absence of dissipation and 
forcing. However, variables in a given shell are influenced 
directly only by their nearest- and next-nearest-neighbour 
shell variables; by contrast, Fourier transformations of the NS and 
the AD equations couple every Fourier mode to every other Fourier 
mode, leading  to the sweeping effect mentioned above. Thus direct 
sweeping is absent in shell models, so they are often thought of as 
an approximate, quasi-Lagrangian representation of their parent 
equations. 

For studies of decaying turbulence one can envisage
several initial conditions. We have used initial conditions of
two types: (a) in the first (Type-I) we drive the system to a 
statistically steady turbulent state by forcing the first shell 
($n = 1$); we then turn off the force and allow the turbulent state 
and the associated energy spectrum to decay freely; our measurements
are made in this decaying state; and (b) for the case of fluid 
turbulence we use a second initial condition (Type-II) in which all 
the energy is concentrated in the first few shells with small $k_n$, 
i.e., large length scales; we then allow the system to evolve 
without any force; the energy cascades to large values of $k_n$ till
the energy spectrum becomes similar to that in forced turbulence; 
this spectrum then decays slowly in time and the measurements we 
report are made during this stage of evolution. 

\subsection{Model B}
\label{modelB}
We use the shell-model analogue of the Kraichnan model introduced 
in~\cite{wirth} in which the equation for the passive-scalar 
variable $\theta_n$ is 
\begin {eqnarray}
\hspace*{-2cm}
\left[\frac{d}{dt} + \kappa k_n^2\right]\theta_n & = & 
\imath\bigg{[}a_n(\theta^*_{n+1}u^*_{n-1} - \theta^*_{n-1}u^*_{n+1}) + 
b_n(\theta^*_{n-1}u^*_{n-2} + 
\theta^*_{n-2}u_{n-1}) \nonumber\\ & + &
c_n(\theta^*_{n+2}u_{n+1} + \theta^*_{n+1}u^*_{n+2})\bigg{]} + f_n,
\label{modelB}
\end{eqnarray}
where the asterisks denote complex conjugation, $a_n = k_n/2$, 
$b_n = -k_{n-1}/2$, and $c_n = k_{n+1}/2$; $f_n$ 
is an additive force that is used to 
drive the system to a steady state; the boundary conditions 
are $u_{-1} = u_0 = \theta_{-1} = \theta_0 = 0; u_{N+1} = u_{N+2} =
\theta_{N+1} = \theta_{N+2} = 0$. The advecting velocity variables 
are taken to be zero-mean, white-in-time, Gaussian random complex 
variables with covariance 
\begin{equation}
\la u_n(t)u_m^*(t^{\prime})\ra = C_2k_n^{-\xi}\delta_{mn}
                        \delta(t - t^\prime),
\label{cor_kshell}
\end{equation}
where $C_2$ is a dimensional constant.
We refer to equations (\ref{modelB}-\ref{cor_kshell}) as Model B. 
 
In our numerical simulations of this model we first obtain a 
statistically steady turbulent state by forcing the first shell with 
a random, Gaussian, white-in-time force. The force is then switched 
off and measurements are made as the turbulence decays. We use a 
weak, order-one, Euler scheme  to integrate the resulting 
Ito form \cite{wirth} of (\ref{modelB}) with an integration time step 
$\delta t = 2^{-24}$, diffusivity $\kappa = 2^{-14}$, and 
$\xi = 0.6$. 

For such a passive-scalar shell model the order-$p$, equal-time, 
structure function and its exponent are defined via 
\begin{equation}
S^{\theta}_p(k_n) \equiv \la [\theta_n(t)\theta^{\ast}_n(t)]^{p/2} 
\ra \sim k_n^{-\zeta^{\theta}_p};
\label{eqstfunB}
\end{equation}
it is natural, therefore, to define the time-dependent version of 
$S^{\theta}_p(k_n)$ as follows:
\begin{equation}
F^{\theta}_p(k_n,t_0,t) \equiv 
\la [\theta_n(t_0)\theta^{\ast}_n(t_0 + t)]^{p/2} \ra.
\label{stfunB}
\end{equation}
The power-law dependence on the right-hand-side of (\ref{eqstfunB}) 
is obtained for $k_n$ in the inertial range. In our numerical 
calculations we use extended self-similarity (ESS) to 
extract the exponent ratios $\zeta_p^\theta/\zeta_2^\theta$. 

\subsection{Model C}
\label{modelC}
The most commonly used shell-model analogue of the NS equation 
is the GOY model~\cite{frisch,gledzer,ohkitani}:  
\begin{eqnarray}
\hspace*{-1cm}
\left[\frac{d}{dt} + \nu k_n^2\right]u_n =  
\imath\bigg{[}a_n u_{n+1}u_{n+2} +   
b_n u_{n-1}u_{n+1} + c_n u_{n-1}u_{n-2} \bigg{]}^{\ast} + f_n. 
\label{goy}
\end{eqnarray}
The coefficients $a_n = k_n$, $b_n = -\delta k_{n-1}$, $c_n = 
-(1-\delta)k_{n-2}$ are chosen in a manner that conserves the 
shell-model analogues of energy and helicity in the inviscid, 
unforced limit; an external force $f_n$ 
drives the system to a steady state. We use the 
standard value $\delta = 1/2$; the boundary conditions are 
$u_{-1} = u_0 = 0; u_{N+1} = u_{N+2} = 0$. We will refer to this as 
Model C.

We use two different kinds of initial conditions in our study of 
decaying fluid turbulence in this model. For Type-I initial 
conditions we first drive the system to a statistically steady 
turbulent state with an external force 
$f_n = (1 + \imath)\times 5\times 10^{-3}\delta_{n,1}$. 
The force is then switched off and the shell velocities at this
instant are taken as the initial condition. The turbulence then 
decays.  Our structure-function measurements are made during this 
period of decay. To obtain the second type of initial condition, the 
energy is initially concentrated in the first few shells by choosing 
the following initial (superscript $0$) velocities: 
$u_n^0 = k_n^{1/2} e^{\imath \vartheta_n},$ for $n = 1, 2$, and 
$u_n^0 = k_n^{1/2}e^{{-k_n}^2} e^{\imath \vartheta_n},$ for 
$3 \leq n \leq N$, with $\vartheta_n$ a random 
phase angle distributed uniformly between $0$ and $2\pi$. 
This energy then cascades down the inertial-range scales without 
significant dissipation until it reaches dissipation-range scales 
at cascade completion. The energy dissipation-rate per unit mass 
shows a peak, as a function of time, roughly 
at cascade completion \cite{chirag,prasad}, and 
the energy spectrum $E(k)$ and the structure function (\ref{eqstfunC}) 
show well-developed inertial ranges.
These decay very slowly in time, so at each instant exponents can 
be determined from plots of $E(k)$ and the structure functions.
Therefore, for initial conditions of this type, 
we wait for cascade completion before making measurements 
of structure functions.  
 
We employ the slaved, Adams-Bashforth scheme \cite{dhar,pisarenko} to 
integrate the GOY-model equations with a time step 
$\delta t = 10^{-4}$. In our numerical simulations the viscosity 
$\nu = 10^{-7}$ and the total number of shells $N = 22$; this 
provides us with a large inertial range from which exponents can be
obtained reliably.

For this model, the order-$p$, equal-time structure function and its 
exponent are defined as follows: 
\begin{equation}
S^{u}_p(k_n) \equiv \la [u_n(t)u^{\ast}_n(t)]^{p/2} \ra 
\sim k_n^{-\zeta^{u}_p};
\label{eqstfunC}
\end{equation}
the associated time-dependent structure function is
\begin{equation}
F^{u}_p(k_n,t_0,t) \equiv 
\la [u_n(t_0)u^{\ast}_n(t_0 + t)]^{p/2} \ra,
\label{goy2}
\end{equation}
where the power-law dependence on the right-hand-side 
of (\ref{eqstfunC}) holds for $k_n$ in the inertial range. (For 
statistically steady turbulence, the time-dependent structure 
function has no dependence on $t_0$, so, without loss of generality, 
$t_0$ can be taken to be 0.) A direct determination of 
$\zeta_p^u$ from (\ref{eqstfunC}) is not very accurate because of an 
underlying 3-cycle in the static version of the GOY shell 
model \cite{kadanoff}. The effects of this 3-cycle can be filtered 
out to a large extent by using the modified structure function 
\begin{equation}
\Sigma^u_p(n) \equiv \la|{\Im}[u_{n+2}u_{n+1}u_n - 
(1/4)u_{n-1}u_nu_{n+1}]|^{p/3} \ra \sim k_n^{-\zeta^{u}_p}; 
\label{sigmau}
\end{equation}
we use $\Sigma^u_p(n)$ in our numerical calculation of 
$\zeta^{u}_p$. We measure time in terms of the initial large 
eddy-turnover time $t_L \equiv 1/(u_{rms}k_1)$; the 
root-mean-square velocity $u_{rms} \equiv [\la \sum_n |u^0_n|^2 \ra]^{1/2}$. 

\subsection{Model D}
\label{modelD}
A turbulent velocity field does not have the  
simple statistical properties assumed in Models A and B.
To overcome this we study the shell model of Ref.\cite{jensen}, 
hereafter referred to as Model D, in which the advecting velocity 
$u_n$ is a solution of the GOY shell model(\ref{goy}). The 
passive-scalar shell variables $\theta_n$ obey 
\begin {eqnarray}
\left[\frac{d}{dt} + \kappa k_n^2\right]\theta_n & = &
\imath\bigg{[}a_n(\theta_{n+1}u_{n-1} - 
\theta_{n-1}u_{n+1}) + 
b_n(\theta_{n-1}u_{n-2} + 
\theta_{n-2}u_{n-1}) \nonumber\\ & + &
c_n(\theta_{n+2}u_{n+1} + \theta_{n+1}u_{n+2})\bigg{]}^* + f_n,
\label{ps2}
\end{eqnarray}
where $a_n = k_n$, $b_n = -k_{n-1}/2$, and 
$c_n = -k_{n+1}/2$; 
$f_n  = (1 + \imath) \times 5 \times 10^{-3}\delta_{n,1}$ is an 
additive force that drives the system to a steady state; the 
boundary conditions are $u_{-1} = u_0 = \theta_{-1} = \theta_0 = 0; 
u_{N+1} = u_{N+2} = \theta_{N+1} = \theta_{N+2} = 0$. 

For this model we start with Type-I initial conditions, i.e., we 
force both the coupled equations (\ref{goy}) and (\ref{ps2}) till a 
statistically steady turbulent state is obtained and then switch off 
the force. The shell variables at this instant of time are taken as 
the initial condition; and then the turbulence is allowed to decay. 

We employ a second-order Adams-Bashforth scheme to integrate Model D 
with a time step $\delta t = 10^{-4}$ and set the diffusivity 
$\kappa = 5 \times 10^{-7}$ so the Schmidt number $\nu/\kappa = 
1/5$; and $N = 22$ as in Model C. The definitions of structure 
functions for Model D are the same as those for Model B, i.e., 
equations (\ref{eqstfunB}) and (\ref{stfunB}). We limit the effects 
of the 3-cycle (mentioned above for Model C) in our numerical 
evaluations of $\zeta_p^\theta$ by using the modified structure 
function
\begin{equation} 
\Sigma_p^\theta(n) \equiv \la 
|{\Im}[\theta_{n+2}\theta_{n+1}\theta_n - 
(1/4)\theta_{n-1}\theta_n\theta_{n+1}]|^{p/3} \ra 
\sim k_n^{-\zeta^{\theta}_p}.
\label{sigmap}
\end{equation}

\section{Analytical Results for Models A and B}
\label{analytic}
In this {\it Section} we present our analytical results for 
time-dependent,  passive-scalar structure functions for the 
Kraichnan model (Model A) 
and its shell-model analogue Model B. For Model A we obtain results 
for both Eulerian and quasi-Lagrangian structure functions. We find, 
in particular, that time-dependent structure functions for these 
models can be factorised into a part that depends on the time origin 
$t_0$ and a part that depends on $t$ but is independent of $t_0$. 
This important result motivates a similar factorisation hypothesis 
that we propose, and verify numerically, for Models C and D in 
subsequent {\it Sections}.  

\subsection{Model A} 

Consider first the Eulerian version of the AD equation 
(\ref{nseqn}). We assume that a turbulent statistical steady state 
has been established because of an external force $f_{\theta}$. We 
turn off this force at time 0. A spatial Fourier transform of 
(\ref{nseqn}) now yields  
\begin{equation}
\frac{\partial\tilde {\theta}({\bf k},t)}{\partial t} = 
\imath\int k_j\tilde{u}_j({\bf q},t)\tilde {\theta}({\bf k} - {\bf q},t)
d^dq -
\kappa k_jk_j\tilde {\theta}({\bf k},t),
\label{advFT}
\end{equation}
where the tildes denote spatial Fourier transforms, we sum over 
repeated indices, and the statistics of $\tilde{u}_j({\bf q},t)$ are 
specified by equations (\ref{covarnce}) and (\ref{Dij1}). 
The Fourier-space, second-order correlation function
${\tilde {\mathcal F}}_2^{\theta}({\bf k},t_0,t) \equiv
 \langle {\tilde \theta}(-{\bf k},t_0){\tilde \theta}({\bf k},t_0 + 
t) \rangle$ ($t_0 \ge 0$ is any time origin) satisfies the equation 
\begin{equation}
\frac{\partial \tilde 
{{\mathcal F}_2^{\theta}}({\bf k},t_0,t)}{\partial t} = 
\langle{\tilde \theta}(-{\bf k},t_0)
\frac{\partial{\tilde \theta}({\bf k},t_0 + t)}
{\partial t}\rangle,
\label{eqnofmotn1}
\end{equation}
which can be combined with (\ref{advFT}) to get 
\begin{eqnarray}
\frac{\partial\tilde 
{{\mathcal F}}_2^{\theta}({\bf k},t_0,t)}{\partial t} &=& 
\imath k_j\int\langle\tilde {\theta}(-{\bf k},t_0)\tilde {u_j}({\bf q},t_0 
+ t)
\tilde {\theta}({\bf k}-{\bf q},t_0 + t)\rangle d^3q \nonumber \\
&-& \kappa k_jk_j\langle\tilde {\theta}(-{\bf k},t_0)
\tilde {\theta}({\bf k},t_0 + t)\rangle.
\label{eqnmotn2}
\end{eqnarray}
We average over the statistics of the advecting velocity field  
(\ref{covarnce}) by using Novikov's theorem \cite{jean}: 
e.g., the first term in (\ref{eqnmotn2}) reduces to 
\begin{eqnarray}
\hspace*{-2cm}
\langle\tilde{\theta}(-{\bf k},t_0)\tilde{u_j}({\bf q},t_0 + t)
\tilde {\theta}({\bf k}-{\bf q},t_0 + t)\rangle = \nonumber\\
\hspace*{-2cm} 
\int_0^{\infty}dt^{\prime}\left[\langle \tilde{u_j}({\bf q},t_0 + t)
\tilde{u_i}(-{\bf q},t^{\prime})\rangle
\langle\tilde {\theta}(-{\bf k},t_0)\frac{\delta}{\delta
\tilde{u_i}(-{\bf q},t_0 + t^{\prime})}
\tilde {\theta}({\bf k}-{\bf q},t_0 + t^{\prime})\rangle\right] .
\end{eqnarray}
Finally equations (\ref{covarnce}) and (\ref{advFT})-(24) yield 
\begin{equation}
\frac{\partial\tilde {{\mathcal F}}_2^{\theta}({\bf k},t_0,t)}
{\partial t} = -2k_ik_j\tilde {{\mathcal F}}_2^{\theta}
({\bf k},t_0,t)\int_0^{\infty}d^dq\tilde{D}_{ij}(q).
\end{equation}
Since $2\int_0^{\infty}{\tilde D}_{ij}d^dq = D^0\delta_{ij} 
\sim C_1L^{\xi}$, where $C_1$ is a dimensional constant, in the limits $\kappa \to 0$, $\eta_d \to 0$, 
and $L \to \infty$ of relevance to turbulence, we get  
\begin{equation}
\frac{\partial {\mathcal F}_2^{\theta}(r,t_0,t)}{\partial t} 
\sim C_1L^{\xi}\frac{\partial^2 {\mathcal F}_2^{\theta}(r,t_0,t)}{\partial r^2}.
\label{E}
\end{equation}
A spatial Fourier transform allows us to integrate this equation to 
obtain 
\begin{equation}
\tilde {{\mathcal F}}_2^{\theta}({\bf k},t_0,t) \sim 
\varphi_2^{\theta}(k,t_0)\exp\left[-C_1L^{\xi}k^2t\right].
\label{K-E}
\end{equation}
If we set $t=0$, we see that 
$\tilde {{\mathcal F}}_2^{\theta}({\bf k},t_0,0)$ is just the 
passive-scalar, equal-time structure function; 
$\varphi_2^{\theta}(k,t_0)$ is, therefore, proportional 
to the Fourier transform of the passive-scalar equal-time structure 
function $\mathcal {S}^{\theta}_2(r)$. Thus, for a fixed but
large value of $L$, we get, in the Eulerian framework, simple 
dynamic scaling with an exponent $z_2^{\cal E} = 2$. We note that 
we get an exponent which is not equal to unity ($\Sref{intro}$) in this 
case because of the white-in-time nature of the advecting field. 
 
The analogue of equation (\ref{E}) for the second-order, 
quasi-Lagrangian structure function follows from \cite{mitra2} the 
quasi-Lagrangian form of the advection-diffusion equation 
(\ref{eq:pas:qlpasscal}). 
\begin{eqnarray}
\frac{\partial {\mathcal F}_2^{\hat \theta}(r,t_0,t)}{\partial t} 
& = &
(D^0\delta_{ij} - D_{ij})\frac{\partial 
{\mathcal F}_2^{\hat \theta}(r,t_0,t)}
{\partial r_i\partial r_j} 
\nonumber \\ 
& \sim & d_{ij}\frac{\partial {\mathcal F}_2^{\hat \theta}(r,t_0,t)} {\partial r_i\partial r_j}. 
\label{QL}
\end{eqnarray}
By substituting for $d_{ij}$ from (\ref{dij}) we obtain, for the 
isotropic case, 
\begin{equation}
\frac{\partial {\mathcal F}_2^{\hat \theta}(r,t_0,t)}
{\partial t} \sim D_1r^{\xi}\frac{\partial^2 {\mathcal F}_2^{\hat 
\theta}(r,t_0,t)}{\partial r^2}. 
\label{QL1}
\end{equation}
A spatial Fourier transformation allows us to integrate this 
equation to obtain, in the limits $\kappa \to 0$, $\eta \to 0$, 
and $L \to \infty$,  
\begin{equation}
\tilde {{\mathcal F}}_2^{\hat \theta}({\bf k},t_0,t) \sim 
\varphi_2^{\hat \theta}(k,t_0)\exp\left[-D_1k^{2-\xi}t\right].
\label{K-QL}
\end{equation}
$\varphi_2^{\hat \theta}(k,t_0)$ is now proportional to the Fourier 
transform of the equal-time, quasi-Lagrangian, passive-scalar 
structure function (which is, of course, the same as the equal-time,
Eulerian, passive-scalar structure function). Equation (\ref{K-QL}) 
shows that, in the quasi-Lagrangian framework, 
$\tilde {{\mathcal F}}_2^{\hat \theta}({\bf k},t_0,t)$ 
factorises into a part that depends on $t_0$ and another which 
depends only on $t$. From the second factor we get simple dynamic 
scaling with an exponent $z_2 = 2 - \xi$, which is different from 
the Eulerian exponent $z_2^{\cal E} = 2$ for Model A. Such a 
factorisation should also follow for higher-order, time-dependent,  
structure functions as we show in the next subsection for Model B. 

\subsection{Model B}
\label{analyticB}
We can use the methods of the previous subsection to obtain 
analytical expressions for time-dependent structure functions for 
Model B.  Consider first the second-order structure function 
\begin{equation}
F^{\theta}_2(n,t_0,t) = 
\langle\theta_n(t_0)\theta^*_n(t_0 + t)\rangle,
\end{equation} 
whence
\begin{equation}
\frac{\partial F^{\theta}_2(n,t_0,t)}{\partial t} = 
\langle\theta_n(t_0)
\frac{\partial\theta^*_n(t_0 + t)}{\partial t}\rangle;
\label{eqnofmotn}
\end{equation}
the angular brackets denote an average over the statistics of 
$u_n(t)$ that are specified by equation (\ref{cor_kshell}). By 
using the complex conjugate of (\ref{modelB}) in (\ref{eqnofmotn}) 
and Novikov's theorem we get terms of the form
\begin{eqnarray}
\hspace*{-2cm}
\langle \theta_n(t_0)\theta^{\ast}_{n+1}(t_0 + t)
u^{\ast}_{n-1}(t_0 + t) \rangle & = &
\langle u^{\ast}_{n-1}(t_0 + t)u_{n-1}(t_0 + t)\rangle\times  
\nonumber \\ 
&&\left\langle\frac{\delta}{\delta {u_{n-1}(t_0 + t)}} 
        \theta_n(t_0) {\theta^{\ast}_{n+1}(t_0 + t)} \right\rangle.
\end{eqnarray}
Finally by using (\ref{cor_kshell}) we obtain
\begin{equation}
\frac{\partial F^{\theta}_2(n,t_0,t)}{\partial t} = 
-\frac{1}{4}C_2k_n^{2-\xi}A(\xi)F^{\theta}_2(n,t_0,t),
\label{}
\end{equation}
where $C_2$ is a dimensional constant.
Integration now yields 
\begin{equation}
F^{\theta}_2(n,t_0,t) = 
\phi^{\theta}_2(n,t_0)\exp\left[-\frac{1}{4}C_2k_n^{2-\xi}A(\xi)
t\right],
\label{decoup}
\end{equation}
with $A(\xi) = (2^{(2\xi - 2)} + 2^{-(2\xi - 2)}) + (2^{\xi} + 
2^{-\xi}) + (2^{(\xi - 2)} + 2^{-(\xi - 2)})$. 
Similarly, we obtain the following exact expression for the 
fourth-order structure function: 
\begin{equation}
F^{\theta}_4(n,t_0,t) = \phi^{\theta}_4(n,t_0)
\exp\left[-\frac{1}{2}C_2k_n^{2-\xi}A(\xi)t\right]. 
\label{fourthordershell}
\end{equation}
$\phi^{\theta}_2(n,t_0)$ and 
$\phi^{\theta}_4(n,t_0)$ are, respectively, the second- and 
fourth-order, equal-time, quasi-Lagrangian, passive-scalar 
structure functions. Thus $z_2 = z_4 = 2 - \xi$, as for the 
quasi-Lagrangian structure functions of Model A. 
(Recall that we expect quasi-Lagrangian behaviour for shell models 
since they do not have a direct sweeping effect.) The equality of
$z_2$ and $z_4$ indicates that we have simple dynamic scaling in 
Model B. We expect all the quasi-Lagrangian exponents $z_p$ to be 
$2-\xi$ for this model, but the analytical demonstration of this 
result becomes more and more complex with increasing $p$.

Given a factorisation of the form shown in (\ref{decoup}), it is 
possible to normalise the time-dependent, passive-scalar structure 
function $F^{\theta}_p(n,t_0,t)$ by its value at $t = 0$ and thus 
make it independent of $t_0$. We cannot prove that such a 
factorisation exists in Models C and D, but we present compelling 
numerical evidence for it in $\Sref{results}$. 

\section{Multifractal Formalism for Models C and D}
\label{multifractal}
Equal-time Eulerian and quasi-Lagrangian structure functions are the 
same for homogeneous and isotropic turbulence~\cite{lebdev}. Since 
quasi-Lagrangian structure functions are required for our study of 
dynamic multiscaling, we present the multifractal formalism in terms 
of quasi-Lagrangian variables. Multiscaling in statistically steady 
fluid turbulence can be rationalised by using the 
multifractal formalism, which assumes that a turbulent flow has a 
continuous set of scaling exponents $h$ in the set 
${\cal I} \equiv (h_{min},h_{max})$, instead of a single exponent 
(e.g., $h=1/3$ yields simple K41 scaling)~\cite{frisch}.  The scaling 
exponents $h$ characterise the behaviour of velocity differences 
$\delta {\hat u}_r({\bf x})$:  For each $h \in {\cal I}$ there 
exists a set ${\bf \Sigma}_h \subset {\mathbb R}^3$ of fractal dimension 
${\cal D}^{\hat u}(h)$ and  
$\delta {\hat u}_r({\bf x})/{\hat u}_L \sim (r/L)^h$ for separations 
$r$ in the inertial range if ${\bf x} \in {\bf \Sigma}_h$ and with ${\hat u}_L$
the velocity at the forcing scale $L$. 
Given the measure $d\mu(h)$ for the weights of the different values of $h$, 
the order-$p$, equal-time velocity structure function is 
\begin{equation}
\frac{{\cal S}^{\hat u}_p(r)}{{\hat u}_L^p} 
\equiv \frac{\la\delta \hat u^p_r({\bf x}) \ra}{{\hat u}_L^p} 
\sim \int_{{\cal I}}{d\mu(h) \big{(}\frac{r}{L}
\big{)}^{ph + 3 - {\cal D}^{\hat u}(h)}},
\label{SPmulti}
\end{equation}
where the $ph$ term comes from $p$ factors of $(r/L)^h$ and the 
additional factor of $(r/L)^{3-{\cal D}^{\hat u}(h)}$ is the 
probability of being within a distance $\sim r$ of the set 
${\bf \Sigma}_h$, of dimension ${\cal D}^{\hat u}(h)$, which is embedded 
in three dimensions.  ${\cal D}^{\hat u}(h)$, $h_{min}$, and 
$h_{max}$ are assumed to be universal. In the limit $r/L \to 0$, of
relevance to fully developed turbulence, we get the equal-time 
scaling exponent 
$\zeta_p^{\hat u} = {\inf}_h[ ph + 3 - {\cal D}^{\hat u}(h)]$
by the method of steepest descents.  

We now define the order-$p$, time-dependent, quasi-Lagrangian 
velocity structure function 
\begin{eqnarray}
{\mathcal F}_p^{\hat u}(r,\{t_1,\ldots,t_p\}) \equiv
        \la [\delta \hat u_{\parallel}({\bf x},r,t_1) \ldots
              \delta \hat u_{\parallel}({\bf x},r,t_p)] \ra. 
\label{Fp}
\end{eqnarray}
For simplicity, we consider $t_1=t$ and $t_2=\ldots=t_p=0$, denote the 
structure function by ${\mathcal F}_p^{\hat u}(r,t)$, and suppress the 
subscript $\parallel$, i.e., we use $\delta \hat u_r({\bf x}) \equiv 
\delta \hat u_{\parallel}({\bf x},r,t_1)$ .  
The natural extension of the multifractal formalism to the 
case of time-dependent structure functions in statistically steady 
fluid turbulence follows from the {\it Ansatz}~\cite{mitra1}
\begin{equation}
\frac{{\cal F}^{\hat u}_p(r,t)}{{\hat u}_L^p} \propto \int_Id\mu(h) 
\left(\frac{r}{L}\right)^{3 + ph 
- {\cal D}^{\hat u}(h)}{\cal G}^{p,h}\left(\frac{t}{\tau_{p,h}}\right),
\label{Fpforced}
\end{equation}
where the scaling function 
${\cal G}^{p,h}\left(\frac{t}{\tau_{p,h}}\right)$ is 
assumed to have a characteristic decay time 
$\tau_{p,h} \sim r/\delta \hat u_r({\bf x}) \sim r^{1-h}$ 
and ${\cal G}^{p,h}(0) = 1$. 

For statistically steady passive-scalar turbulence the application 
of the multifractal formalism is more complicated than it is for 
fluid turbulence. This is because we must now deal with a joint 
multifractal distribution of both the velocity and passive-scalar 
variables. Therefore, the order-$(p,p^{\prime})$, equal-time 
structure function for passive-scalar turbulence 
\begin{equation}
{\mathcal S}_{p,p^{\prime}}^{{\hat \theta},{\hat u}}(r) \equiv  
\langle \delta {\hat \theta}^p_r({\bf x})\delta 
{\hat u}^{p^{\prime}}_r({\bf x})\rangle
\label{jointsp}
\end{equation}
has the multifractal representation
\begin{equation}
\frac{{\mathcal S}^{{\hat \theta},{\hat u}}_{p,p^{\prime}}(r)}
{{\hat \theta}_L^p {\hat u}_L^{p^{\prime}}}
\propto \int_{{\cal I},{\cal I}^{\prime}}d\mu(h,g)
\left(\frac{r}{L}\right)^{3+pg+p^{\prime}h -
{\cal D}^{{\hat u},{\hat \theta}}(h,g)},
\end{equation}
where ${\hat u}$ and ${\hat \theta}$ are assumed to possess a range 
of universal scaling exponents $h \in{\cal I} 
\equiv (h_{min},h_{max})$ and $g\in{\cal I}^{\prime} 
\equiv (g_{min},g_{max})$, respectively.
For each pair of $h$ and $g$ in these ranges, there exists a set
${\bf \Sigma}_{h,g} \subset \mathbb{R}^3$ of fractal dimension 
${\cal D}^{{\hat u},{\hat \theta}}(h,g)$. The increments in the 
velocity $\delta{\hat u}_r({\bf x})$ and the passive-scalar field 
$\delta{\hat\theta}_r({{\bf x}})$ scale as 
$\delta{\hat u}_r({\bf x})/{\hat u}_L \sim (r/L)^h $ and 
$\delta{\hat\theta}_r({{\bf x}})/{\hat \theta}_L \sim (r/L)^g$ for 
separations $r$ in the inertial range if 
${\bf x} \in {\bf \Sigma}_{h,g}$. ${\hat u}_L$ and ${\hat \theta}_L$ 
are, respectively, the velocity and the passive-scalar variables at the 
forcing scale $L$. For simplicity we will only consider 
passive-scalar structure functions with $p^{\prime} = 0$ in 
equation (\ref{jointsp}), i.e., 
\begin{equation}
{\mathcal S}_p^{{\hat \theta}}(r) \equiv  
\langle \delta {\hat \theta}^p_r({\bf x})\rangle,
\end{equation} 
which have the multifractal representation 
\begin{equation}
\frac{{\mathcal S}^{{\hat \theta}}_{p}(r)}{{\hat \theta}_L^p}
\propto \int_{{\cal I} {\cal I}^{\prime}}d\mu(h,g)
\left(\frac{r}{L}\right)^{3+pg-{\cal D}^{{\hat u},{\hat \theta}}(h,g)};
\end{equation}
as before, the equal-time exponents $\zeta_p^{\hat \theta}$ can be 
related to ${\cal D}^{{\hat u},{\hat \theta}}(h,g)$ by the method of 
steepest descents. The corresponding order-$p$, time-dependent 
passive-scalar structure function is 
\begin{equation}
{\mathcal F}_p^{\hat {\theta}}(r,\{t_1,\ldots,t_p\}) \equiv 
\langle[\delta\hat\theta({\bf x},r,t_1)...
\delta\hat\theta({\bf x},r,t_p)]\rangle.
\label{Fp1}
\end{equation}
As in (\ref{Fp}), we consider $t_1=t$ and $t_2=\ldots=t_p=0$
for simplicity. Given the nature of multiscaling in passive scalars 
advected by a turbulent velocity field, it is possible to understand 
passive-scalar turbulence within the framework of the multifractal 
formalism. However, the analogous expression for time-dependent 
structure functions in passive-scalar turbulence has to take into 
account the multifractal nature of the advecting velocity 
field \cite{mitra2}. We generalise the multifractal representation of 
time-dependent structure functions in the following way:
\begin{equation}
\frac{{\mathcal F}^{{\hat \theta}}_p(r,t)}{{\hat \theta}_L^p}\propto \int_{{\cal I}, {\cal I}^{\prime}}d\mu(h,g)
\left(\frac{r}{L}\right)^{3+pg-{\cal D}^{{\hat u},{\hat \theta}}(h,g)}
{\cal G}^{p,h,g}\left(\frac{t}{\tau_{phg}}\right);
\label{multips1}
\end{equation}
and we assume that (a) the function 
${\cal G}^{p,h,g}(\frac{t}{\tau_{phg}})$ 
has a characteristic decay time $\tau_{phg}$, (b) 
${\cal G}^{p,h,g}(0) = 1$, and (c) that the dominant 
contribution to $\tau_{phg}$, in the limits  $\kappa \rightarrow 0$,
$\eta_d \rightarrow 0$ and $L \rightarrow \infty$, has the following 
scaling form 
\begin{equation}
\tau_{phg} \sim r/\delta {\hat u}_r({\bf x}) \sim r^{1-h}.
\label{tau}
\end{equation}

Given ${\mathcal F}^{\hat u}_p(r,t)$ and 
${\mathcal F}^{\hat\theta}_p(r,t)$, we define the order-$p$, 
degree-$M$, integral-($I$) and derivative-($D$) times as 
follows \cite{mitra1,mitra2} (if the integrals and derivatives in equations 
(\ref{timp}) and (\ref{tdp}) exist): 
\begin{equation}
{\cal T}^{I,\hat \phi}_{p,M}(r) \equiv
 \left[ \frac{1}{{\mathcal S}^{\hat \phi}_p(r)}
\int_0^{\infty}{\mathcal F}^{\hat \phi}_p(r,t)t^{(M-1)} dt
\right]^{(1/M)}
\label{timp}
\end{equation}
and
\begin{equation}
{\cal T}^{D,\hat \phi}_{p,M}(r) \equiv
 \left[ \frac{1}{{\mathcal S}^{\hat \phi}_p(r)}
\frac{\partial^M {\mathcal F}^{\hat \phi}_p(r,t)}{\partial t^M}
\biggl{|}_{t=0} \right]^{(-1/M)},
\label{tdp}
\end{equation}
respectively, with $\hat \phi$ either $\hat u$ or $\hat\theta$. 
For statistically steady turbulence we can then use the 
dynamic-multiscaling {\it Ansatz} to define the integral- and 
derivative-time multiscaling exponents for fluid turbulence 
$z^{I,{\hat u}}_{p,M}$ and $z^{D,{\hat u}}_{p,M}$  via 
\begin{equation}
{\cal T}^{I,{\hat u}}_{p,M}(r) \sim r^{z^{I,{\hat u}}_{p,M}}
\label{Inttime}
\end{equation}
and
\begin{equation}
{\cal T}^{D,{\hat u}}_{p,M}(r) \sim r^{z^{D,{\hat u}}_{p,M}},
\label{Dertime}
\end{equation} 
respectively, for $r$ in the inertial range.

By substituting the multifractal form (\ref{Fpforced}) in 
(\ref{Inttime}), evaluating the time integral first, and then 
performing the integration over the multifractal measure by the 
saddle-point method, we obtain the {\it integral} bridge relations
\begin{equation}
z^{I,{\hat u}}_{p,M} = 1 + \left[\zeta^{{\hat u}}_{p-M} - 
\zeta^{{\hat u}}_p\right]/M, 
\label{zipm}
\end{equation}
which were first obtained in \cite{lvov}.  Similarly we get the 
{\it derivative} bridge relations
\begin{equation}
z^{D,{\hat u}}_{p,M} = 1 + \left[\zeta^{{\hat u}}_p - 
\zeta^{{\hat u}}_{p+M}\right]/M,
\label{zdpm}
\end{equation}
which were first obtained in forced Burgers 
turbulence \cite{hayot1,hayot2} for the special cases 
(a) $p = 2$, $M = 1$ and (b) $p = 2$, $M = 2$, 
respectively, and for statistically steady fluid turbulence 
in \cite{mitra1}. We note that within the K41 phenomenology 
($\zeta_p^{u,K41} = p/3$) the bridge relations yield the same
dynamic exponent $z_p^{K41} = 2/3$ for both integral- and 
derivative-time scales.

For passive-scalars advected by a turbulent velocity field, the 
corresponding dynamic-multiscaling exponents are defined via
\begin{equation} 
{\cal T}^{I,{\hat \theta}}_{p,M}(r) 
\sim r^{z^{I,{\hat \theta}}_{p,M}}
\label{Inttime1}
\end{equation}
 and 
\begin{equation}
{\cal T}^{D,{\hat \theta}}_{p,M}(r) 
\sim r^{z^{D,{\hat \theta}}_{p,M}}.
\label{Dertime1}
\end{equation}
To obtain bridge relations for dynamic-multiscaling exponents 
in statistically steady passive-scalar turbulence we define 
the degree-$M$, order-$p$ {\it integral}-time for this case:
\begin{eqnarray}
\hspace*{-2.0cm}
{\cal T}^{I,{\hat \theta}}_{p,1}(r) \equiv \left[
\frac{1}{{\mathcal S}^{{\hat \theta}}_p(r)}
\int_{I,I^{\prime}}d\mu(h,g)\left(\frac{r}{L}\right)^{Z}
\tau_{p,g,h}^M 
\int_0^{\infty}{\cal G}^{p,h,g}
\left(\frac{t}{\tau_{pgh}}\right)^{M-1}d\left(\frac{t}{\tau_{pgh}}\right)\right]^{1/M},
\end{eqnarray}
where $Z = 3 + pg - {\cal D}^{{\hat u},{\hat \theta}}(h,g)$ and the 
argument of the scaling function is suppressed for notational 
convenience.  By using the scaling form of $\tau_{pgh}$ (\ref{tau})and 
assuming, as in Ref.\cite{jensen}, that
\begin{equation}
\langle \delta\hat\theta^p\delta \hat u^{-q} \rangle
  \approx \langle\delta\hat\theta^p \rangle
          \langle \delta \hat u^{-q} \rangle,
\label{assump}
\end{equation}
we get 
\begin{equation}
{\cal T}^{I,{\hat \theta}}_{p,1}(r)
\sim \biggl[ \frac{r^M}{{\mathcal S}^{{\hat \theta}}_p(r)}
\langle \delta\hat\theta^p \rangle 
\langle \delta \hat u^{-M} \rangle\biggl]^{1/M}.  
\end{equation}
Thus the degree-$M$, order-$p$, integral-time, dynamic-multiscaling 
exponent $z^{I,{\hat \theta}}_{p,1} = 1 + \zeta^{\hat u}_{-M}/M$.

Similarly we obtain the degree-$M$, order-$p$ {\it derivative} time 
scale for passive-scalars by substituting for 
${\cal F}_p^{{\hat \theta}}$ from (\ref{multips1}) in (\ref{tdp}) and 
using (\ref{tau}) and (\ref{assump}):
\begin{eqnarray}
{\cal T}^{D,{\hat \theta}}_{p,M}(r) & = & \left[ 
\frac{1}{{\mathcal S}^{\hat \theta}_p(r)}\int_{I,I^{\prime}}d\mu(h,g)
\left(\frac{r}{L}\right)^{Z}\left(\frac{\partial^M}{\partial t^M}
{\cal G}^{p,h,g}\biggl{|}_{t=0}\right)\frac{1}{\tau^M_{pgh}}\right]^{-1/M}
\nonumber
\\
& \sim &
\left[ \frac{1}{{\mathcal S}^{\hat \theta}_p(r)}
\int_{I,I^{\prime}}d\mu(h,g)\left(\frac{r}{L}\right)^{Z}
\left(\frac{\partial^M}{\partial t^M}{\cal G}^{p,h,g}\biggl{|}_{t=0}\right)
\frac{[\delta \hat u(r)]^M}{r^M}\right]^{-1/M}
\nonumber
\\
& \sim &
\left[ \frac{1}{r^M{\mathcal S}^{\hat \theta}_p(r)}\la 
\delta \hat u^M(r) 
\delta \hat \theta^p(r) \ra\right]^{-1/M}
\label{d1}
\nonumber
\\
& \sim & 
\left[\frac{1}{r^M{\mathcal S}^{\hat \theta}_p(r)}\la 
\delta \hat u^M(r) \ra 
\la \delta \hat \theta^p(r) \ra\right]^{-1/M} 
\nonumber 
\\
& \sim &  r^{1 - \zeta^{\hat u}_M/M}, 
\end{eqnarray}
which, along with (\ref{Dertime1}), yields the bridge relation  
$z^{D,{\hat \theta}}_{p,M} = 1 - \zeta^{\hat u}_{M}/M$.
To summarise, the bridge-relations for degree-$M$, order-$p$, 
derivative- and integral-time dynamic-multiscaling exponents are, 
respectively:
\begin{eqnarray}
z^{I,{\hat \theta}}_{p,M} = 1+\frac{\zeta^{\hat u}_{-M}}{M}; \hspace*{0.4cm}
z^{D,{\hat \theta}}_{p,M} = 1-\frac{\zeta^{\hat u}_M}{M}.
\label{eq:pas:zp_pshell}
\end{eqnarray}
Note that, in contrast to the bridge relations (\ref{zipm}) and (\ref{zdpm}), 
there is no $p$-dependence on the right-hand sides of these relations. However 
the $M$-dependence is the signature of nontrivial dynamic multiscaling here. 
Furthermore, the integral scale bridge relation is meaningful only for those 
values of $M$ for which $\zeta^{\hat u}_{-M}$ is well defined ~\cite{mitra2}.
We now extend our discussion to the the case of decaying turbulence 
and define the order-$p$, time-dependent quasi-Lagrangian velocity 
structure function (for fluid turbulence) and passive-scalar 
structure function (for passive-scalar turbulence) as 
\begin{eqnarray}
{\mathcal F}_p^{\hat u}(r,\{t_1,\ldots,t_p\}) \equiv
        \la [\delta \hat u_{\parallel}({\bf x},r,t_1) \ldots
              \delta \hat u_{\parallel}({\bf x},r,t_p)] \ra 
\label{Fp3}
\end{eqnarray}
and
\begin{equation}
{\mathcal F}_p^{\hat {\theta}}(r,\{t_1,\ldots,t_p\}) \equiv 
\langle[\delta\hat\theta({\bf x},r,t_1)...
\delta\hat\theta({\bf x},r,t_p)]\rangle,
\label{Fp4}
\end{equation}
respectively. For simplicity, we consider the case $t_1=t_0 + t$ 
and $t_2=\ldots=t_p=t_0$. The derivations of the bridge relations 
we have given above go through if we {\it assume} the following 
multifractal forms for time-dependent velocity and passive-scalar 
structure functions, respectively, in decaying turbulence:
\begin{equation}
\frac{{\cal F}^{\hat u}_p(r,t_0,t)}{{\hat u}_L^p} \propto 
A^{\hat u}(r,t_0)\int_{\cal I}d\mu(h) 
\left(\frac{r}{L}\right)^{3 + ph - {\cal D}^{\hat u}(h)} 
{\cal G}^{p,h}\left(\frac{t}{\tau_{p,h}}\right);
\label{Fpdecay}
\end{equation}
and 
\begin{equation}
\frac{{\mathcal F}^{{\hat \theta}}_p(r,t_0,t)}{{\hat \theta}_L^p}
\propto A^{\hat \theta}(r,t_0) \int_{{\cal I}, 
{\cal I}^{\prime}}d\mu(h,g)
\left(\frac{r}{L}\right)^{3+pg-{\cal D}^{{\hat u}, 
{\hat \theta}}(h,g)}{\cal G}^{p,h,g}\left(\frac{t}{\tau_{pgh}}\right);
\label{multipsd}
\end{equation}
i.e., we assume that the multifractal form factorises into a part
$A^{\hat u}(r,t_0)$ [or $A^{\hat \theta}(r,t_0)$ in the case of 
passive scalars], which depends on the origin of time $t_0$,
and an integral that is independent of $t_0$. This assumption 
is motivated by the factorisation we have seen in $\Sref{analytic}$ above in 
(\ref{K-QL}), (\ref{decoup}), and (\ref{fourthordershell}) for time-dependent 
passive-scalar structure functions in Models A and B. [The 
analogous expressions for equal-time structure functions in 
decaying turbulence are obtained by setting $t=0$ in 
(\ref{Fpdecay}) and (\ref{multipsd}).]  
Given (\ref{Fpdecay}) and (\ref{multipsd}), the integral- and 
derivative-time scales for decaying turbulence become 
{\it independent} of $t_0$ and assume forms identical 
to (\ref{timp}) and (\ref{tdp}); consequently the bridge 
relations (\ref{zipm}), (\ref{zdpm}) and (\ref{eq:pas:zp_pshell}) 
remain unaltered in decaying turbulence.

In the next {\it Section} we give compelling numerical evidence in 
support of our assumptions (\ref{Fpdecay}) and (\ref{multipsd}).

\section{Numerical Results}
\label{results}
We now present our numerical results for dynamic multiscaling in 
decaying, homogeneous, isotropic turbulence in Models B,C, and D 
in $\Sref{b}$, $\Sref{c}$, and $\Sref{d}$, respectively. We show {\it 
en passant} that, for Models B and D, equal-time multiscaling 
exponents are universal in the sense that the exponents obtained 
from decaying-turbulence runs are equal (within error bars) to the 
exponents for statistically steady turbulence in these 
models \cite{mitra1, wirth, jensen}. The underlying reason for
this universality for passive-scalar turbulence has been
discussed earlier \cite{arad,cohen}, but we believe that ours is
the first numerical demonstration of such universality in these 
models. Since we use shell models in this {\it Section}, we replace 
$\hat u$ and $\hat \theta$ by $u$ and $\theta$, respectively.	

\subsection{Model B}
\label{b}
\begin{figure}[htbp]
\begin{center}
\includegraphics[height=6cm,width=6cm]{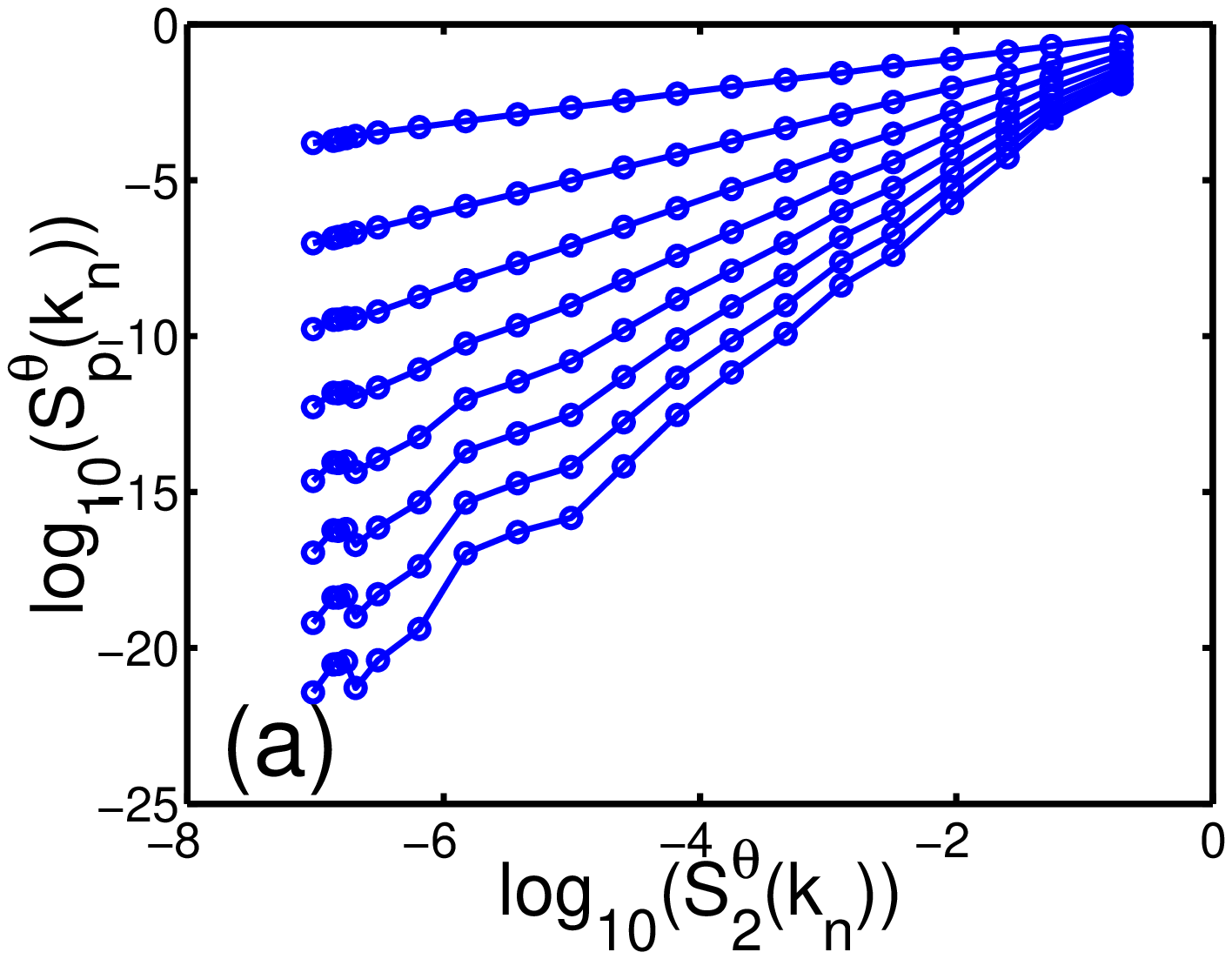}
\includegraphics[height=6cm,width=6cm]{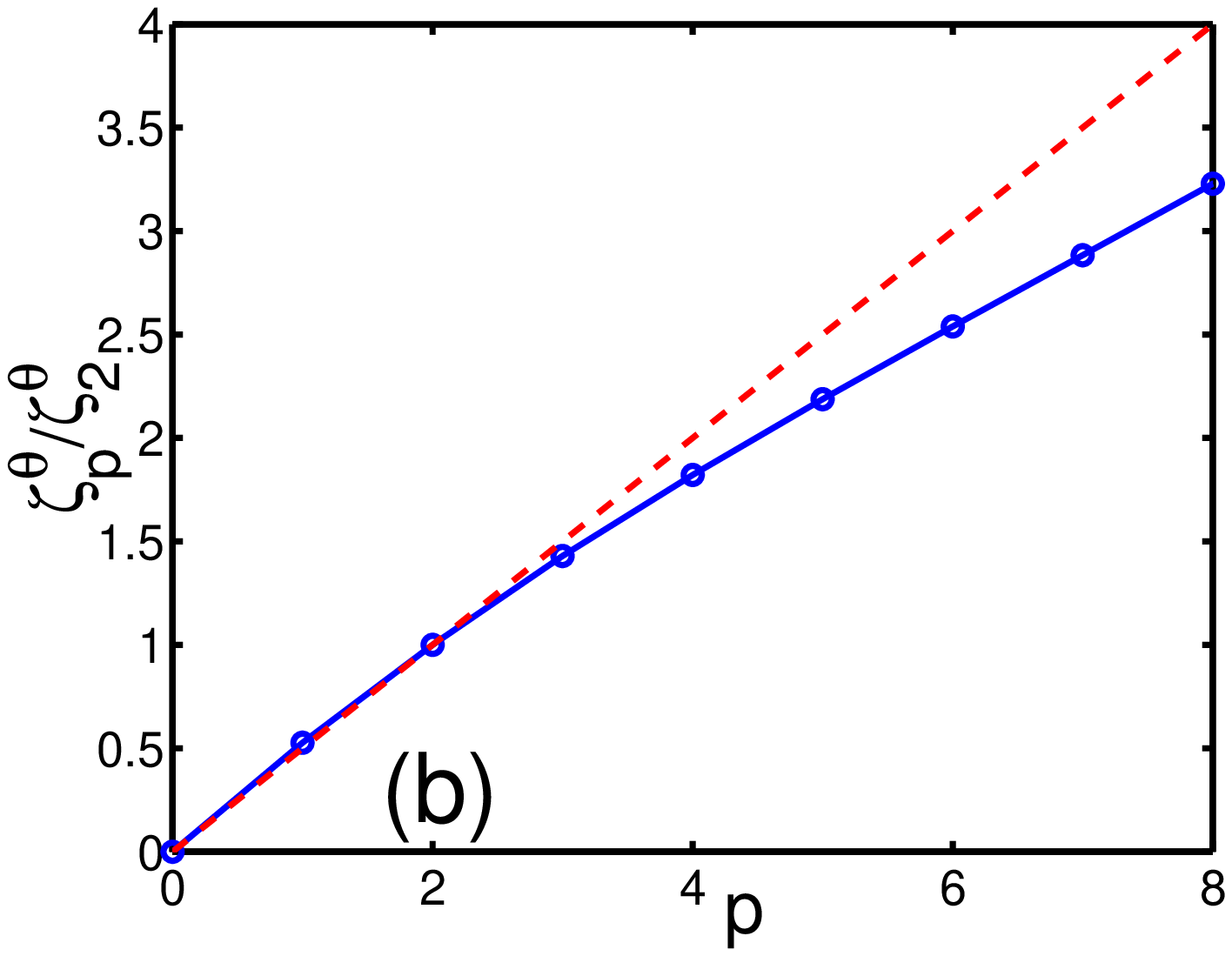}  
\caption{(a) Log-log plots of the order-$p$ structure functions
 $S_p^{\theta}(k_n)$ versus the second-order structure function 
$S_2^{\theta}(k_n)$ from our numerical simulations of Model B for 
$p=1$ (uppermost curve) to $p=8$ (lowermost curve). (b) Plot 
of the exponent ratios $\zeta^{\theta}_p/\zeta^{\theta}_2$, obtained
from the slopes of the linear, inertial region in (a), versus $p$. 
Our data points ($\circ$) are connected by a line to guide
the eye; the error bars are smaller than the sizes of our symbols; 
the dashed straight line corresponds to the analogue of the K41 
scaling prediction for this case.}
\label{statB}
\end{center}
\end{figure}

Let us begin with the equal-time exponents $\zeta_p^{\theta}$  
for the case of decaying turbulence in Model B. As in Ref. 
\cite{wirth}, we use the extended-self-similarity (ESS) procedure
to obtain exponent ratios from log-log plots [Fig. 
(\ref{statB}a)] of the equal-time structure functions 
$S_p^{\theta}(k_n)$ versus $S_2^{\theta}(k_n)$. In 
particular, the slopes of the linear regions of the plots in Fig. 
(\ref{statB}a), with $4\leq n \leq 12$, yield the 
inertial-range exponent ratios $\zeta_p^{\theta}/\zeta_2^{\theta}$.  
From 50 such statistically independent runs (each run, in turn, is 
averaged over 5000 independent initial conditions) we calculate the 
means of the equal-time multiscaling exponent ratios 
$\zeta_p^{\theta}/\zeta_2^{\theta}$; these are plotted in 
Fig. (\ref{statB}b) as functions of the order $p$ for 
$1\leq p \leq 8$. The standard deviations of these exponent ratios 
(calculated from our 50 runs) provide us with error bars; these
are smaller than the symbol sizes used in Fig. (\ref{statB}b). 
Our exponents ratios are in agreement (within the error bars) with 
earlier results for equal-time exponents for statistically steady  
passive-scalar turbulence in Model B \cite{mitra2,wirth}. We have 
also determined the exponent $\zeta_2^{\theta}$ directly from 
log-log plots of $\Sigma_2^{\theta}(k_n)$ versus $k_n$ [equation (\ref{sigmap})]. 
We find  $\zeta_2^{\theta} = 1.403 \pm 0.003$; this agrees well with the 
analytical prediction \cite{wirth,kraich4} 
$\zeta_2^{\theta} = 2 - \xi$, since we use $\xi = 0.6$ in 
our simulations.

\begin{figure}[htbp]
\begin{center}
\includegraphics[height=6cm,width=6cm]{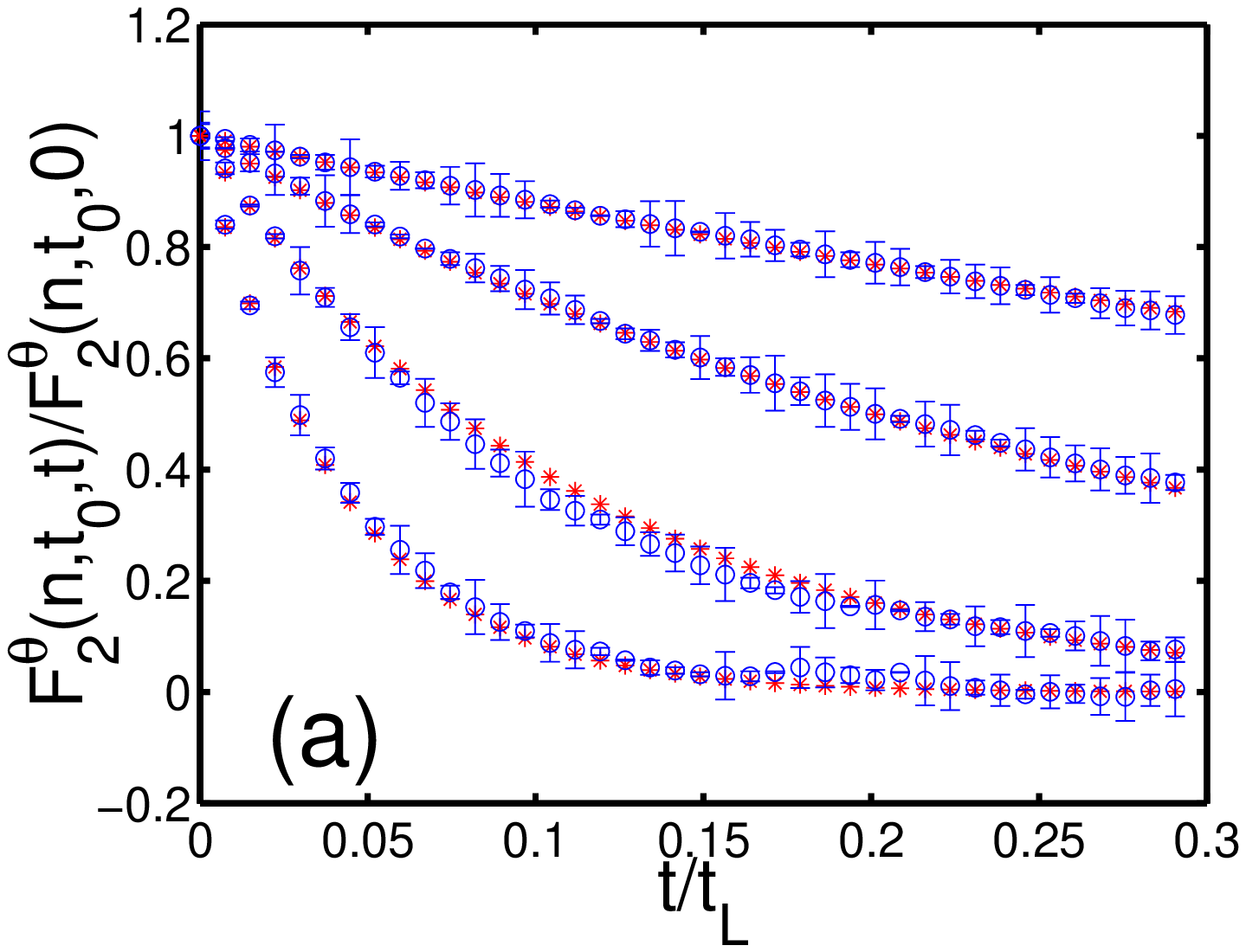}
\includegraphics[height=6cm,width=6cm]{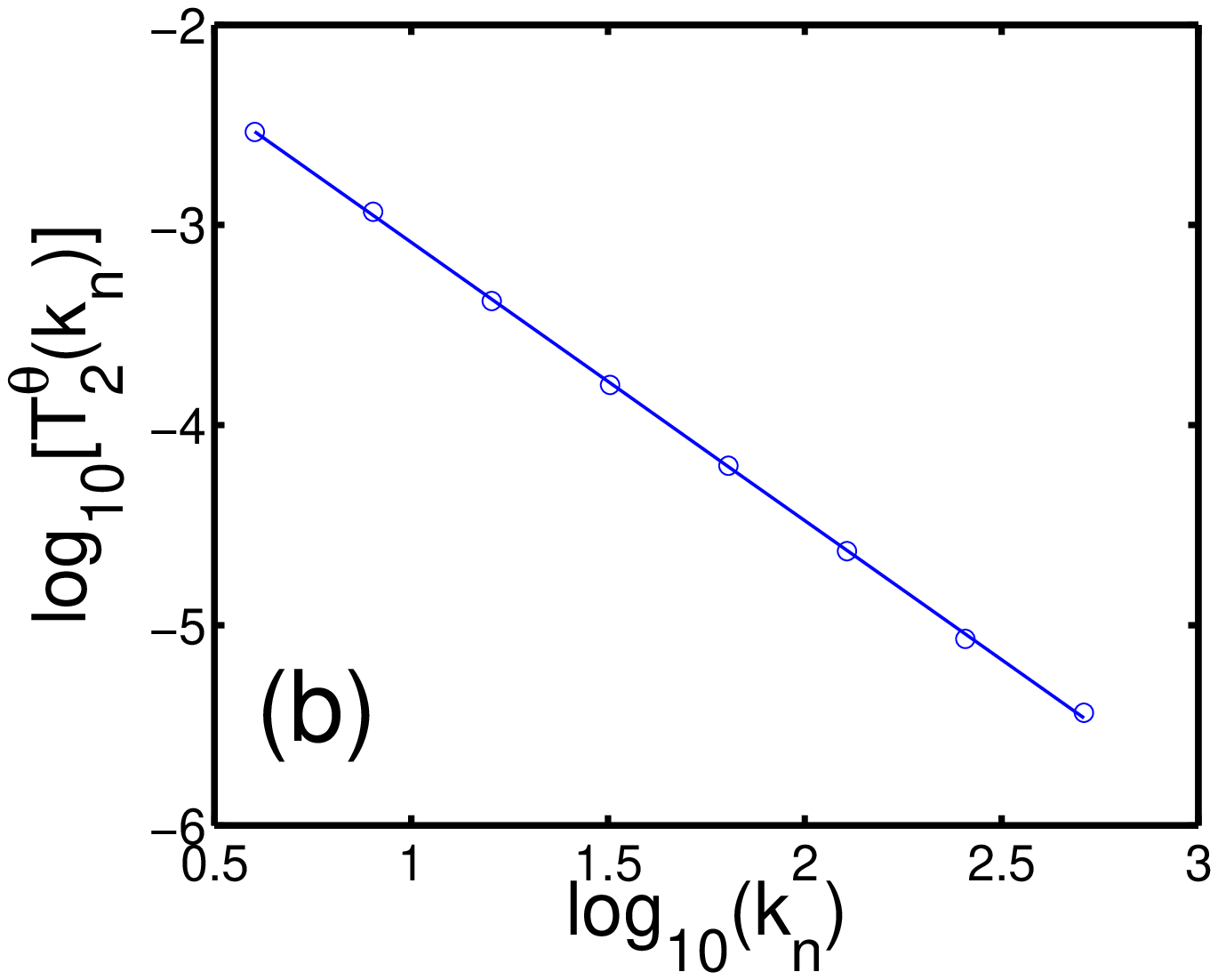}
\caption{(a) Plots comparing the normalised, second-order, 
time-dependent structure function versus the dimensionless time $t/t_L$ 
from our numerical simulations (o) of Model B with our analytical 
expression (\ref{decoup}) (*). For clarity we show data for four shells,
 from $n = 6$ (uppermost) to $n = 9$ (lowermost). 
The error bars on our numerical simulations are shown as vertical lines 
on the data points.
\\(b) A log-log plot of $T^{\theta}_2(k_n)$ versus $k_n$. The slope of this 
plot gives the dynamic scaling exponent $z_2 = 1.398 \pm 0.003$. 
Our result from numerical simulations agree well with the analytical 
result $z_2 = 2 - \xi$, since we have chosen $\xi = 0.6$.}
\label{fit}
\end{center}
\end{figure}

\begin{figure}[htbp]
\begin{center}
\includegraphics[height=6cm,width=6cm]{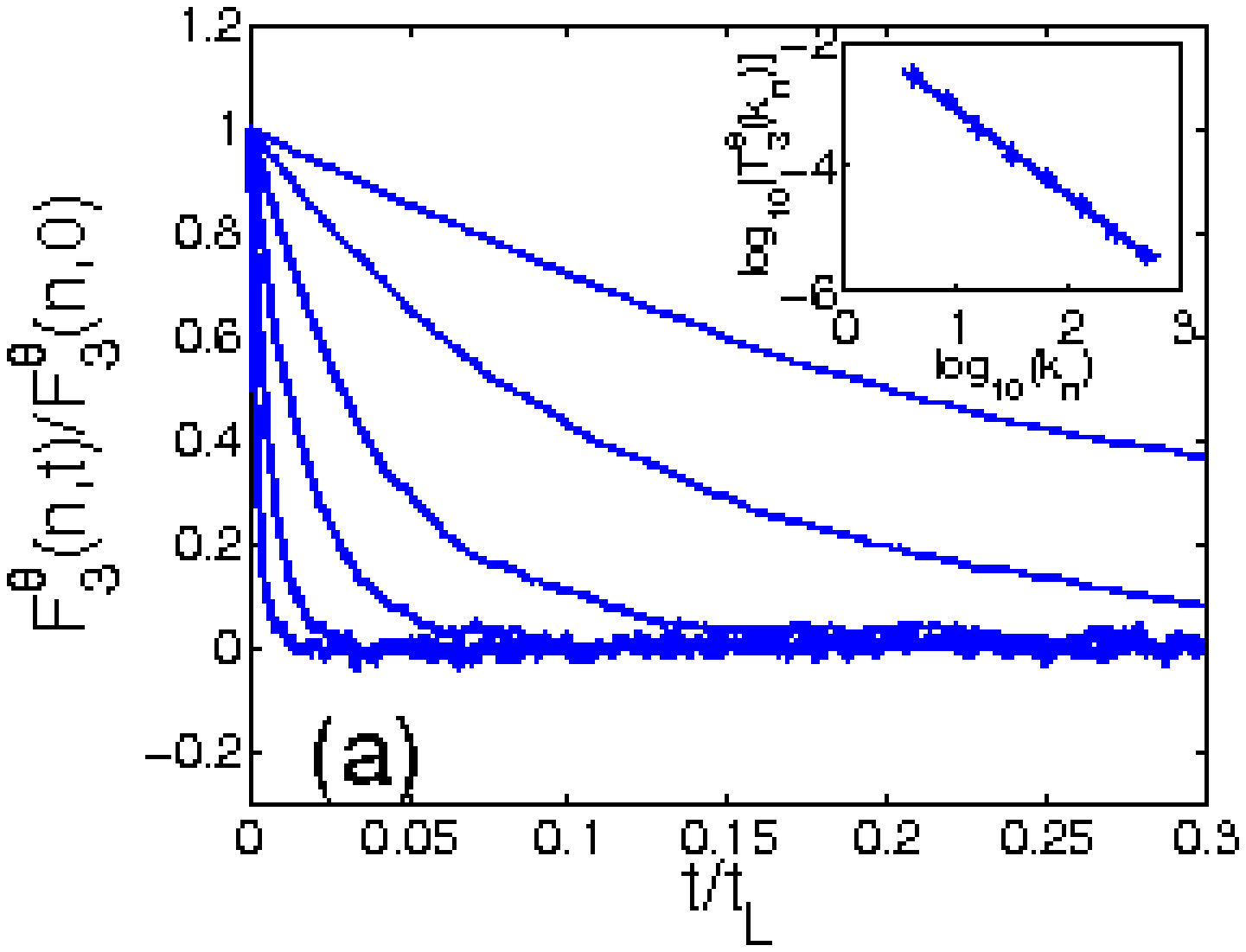}
\includegraphics[height=6cm,width=6cm]{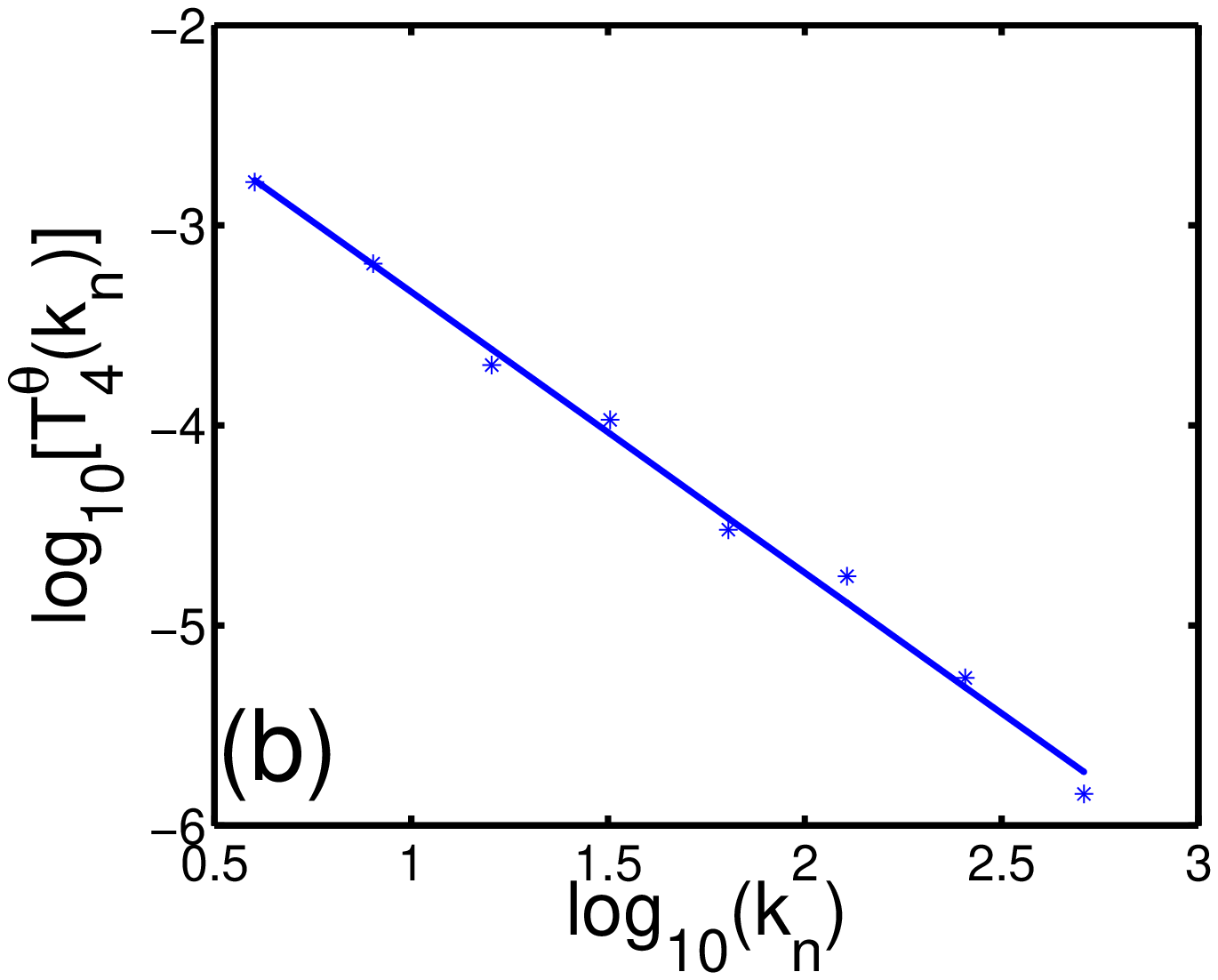}
\caption{(a) Representative plots of the normalised, third-order time-dependent structure 
function versus the dimensionless time $t/t_L$ from our numerical 
simulations of Model B for shell numbers $n = 6$ (uppermost) to $n = 11$ 
(lowermost).  The inset shows a log-log 
plot of $T^{\theta}_3(k_n)$ versus $k_n$. The slope of this plot gives 
the dynamic scaling exponent $z_3 = 1.398 \pm 0.003$, which agrees with $z_2$ 
from Fig. 2b. 
\\(b) A log-log plot of $T^{\theta}_4(k_n)$ versus $k_n$. The slope of this 
plot gives the time-dependent scaling exponent $z_4 = 1.402 \pm 0.005$. 
Our result from numerical simulations agree well with the analytical 
result $z_2 = z_4 = 2 - \xi$, since we have chosen $\xi = 0.6$.}
\label{fit34}
\end{center}
\end{figure}

We begin the discussion of our numerical results for time-dependent, 
passive-scalar structure functions in decaying turbulence by 
comparing our analytical result (\ref{decoup}) with data from our 
simulations of Model B. In Fig. (\ref{fit}a) we show the agreement 
between the two. The error bars have been calculated as in the 
the case of equal-time multiscaling exponents: The data we present
are the means of the values obtained from 50 different statistically
independent runs; and the standard deviations of these values yield 
the error bars. Here and henceforth time is scaled by the 
large-eddy-turnover time $t_L$. Given the exponential decays of 
time-dependent passive-scalar structure functions in Models A and B
[equations (\ref{K-QL}) and (\ref{decoup})], we can extract a 
{\it unique} time scale from plots like the one in Fig. 
(\ref{fit}b). Thus we do not expect time-dependent structure 
functions to exhibit dynamic multiscaling in Models A and B. 
In particular,  equation (\ref{decoup}) implies 
$T^{\theta}_2(k_n) = [1/4k_n^{2-\xi}A(\xi)]^{-1}$ whence 
$z_2 = 2-\xi.$ For our choice of $\xi = 0.6$ we should, therefore, 
obtain $z_2 = 1.4$, and indeed our numerical simulations yield 
$z_2 = 1.398 \pm 0.003$ [Fig. (\ref{fit}b)]. To demonstrate 
numerically that higher-order time-dependent structure functions 
also have a dynamic scaling exponent equal to $z_2$, as shown in 
$\Sref{analyticB}$ for the fourth-order time-dependent structure
function, we show in Figs. (\ref{fit34}a) and (\ref{fit34}b) our numerical results for 
the third- and fourth-order time-dependent structure functions 
versus $t/t_L$ for $6\leq n \leq 13$ from which we extract 
$T^{\theta}_3(k_n)$ and $T^{\theta}_4(k_n)$, respectively. 
The insets show log-log plots of these times versus 
$k_n$; the slopes of these plots yield the dynamic-scaling 
exponents $z_3 = 1.398 \pm 0.003$ and $z_4 = 1.402 
\pm 0.005$ in agreement with our expectation $z_p = 2 - \xi$,
for all $p$, since $\xi = 0.6$ in our calculations.
 
\subsection{Model C}
\label{c}
The equality of equal-time exponents for decaying and statistically 
steady fluid turbulence was demonstrated numerically for the Sabra 
shell model in Ref.~\cite{lvov1}. We have carried out a 
similar exercise for the GOY shell model of fluid turbulence and 
found, unsurprisingly, that the universality of equal-time 
multiscaling exponents holds for the GOY model as well. 
Furthermore, we have also checked that a replacement of the viscous 
term $\nu_0k_n^2$ in equation (\ref{goy}) by the hyperviscous term 
$\nu_{\alpha} k_n^2(k_n/k_d)^{\alpha}$ does not affect these
exponents~\cite{sp23}.

\begin{figure}[htbp]
\begin{center}
\includegraphics[height=6cm,width=6cm]{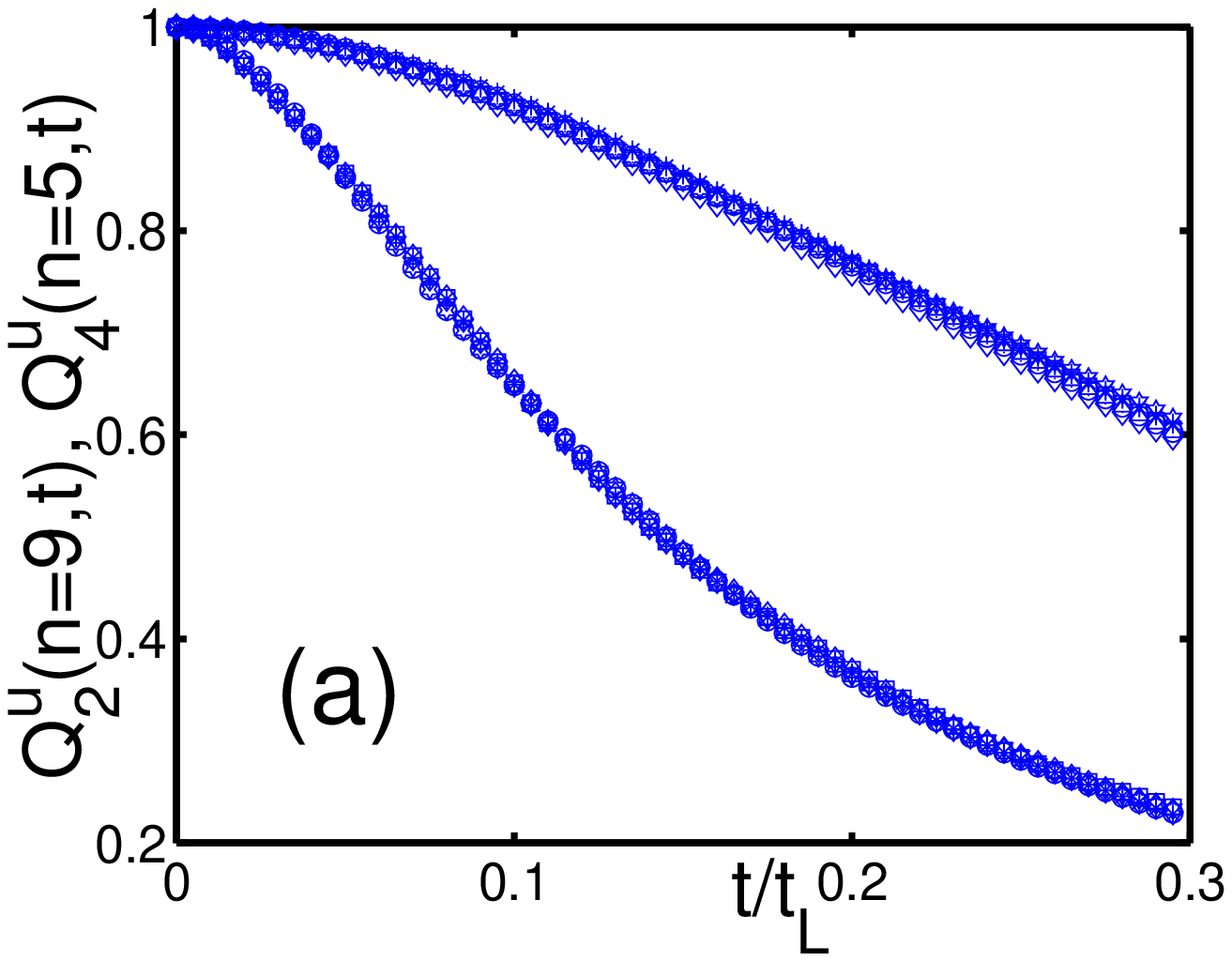} 
\includegraphics[height=6cm,width=6cm]{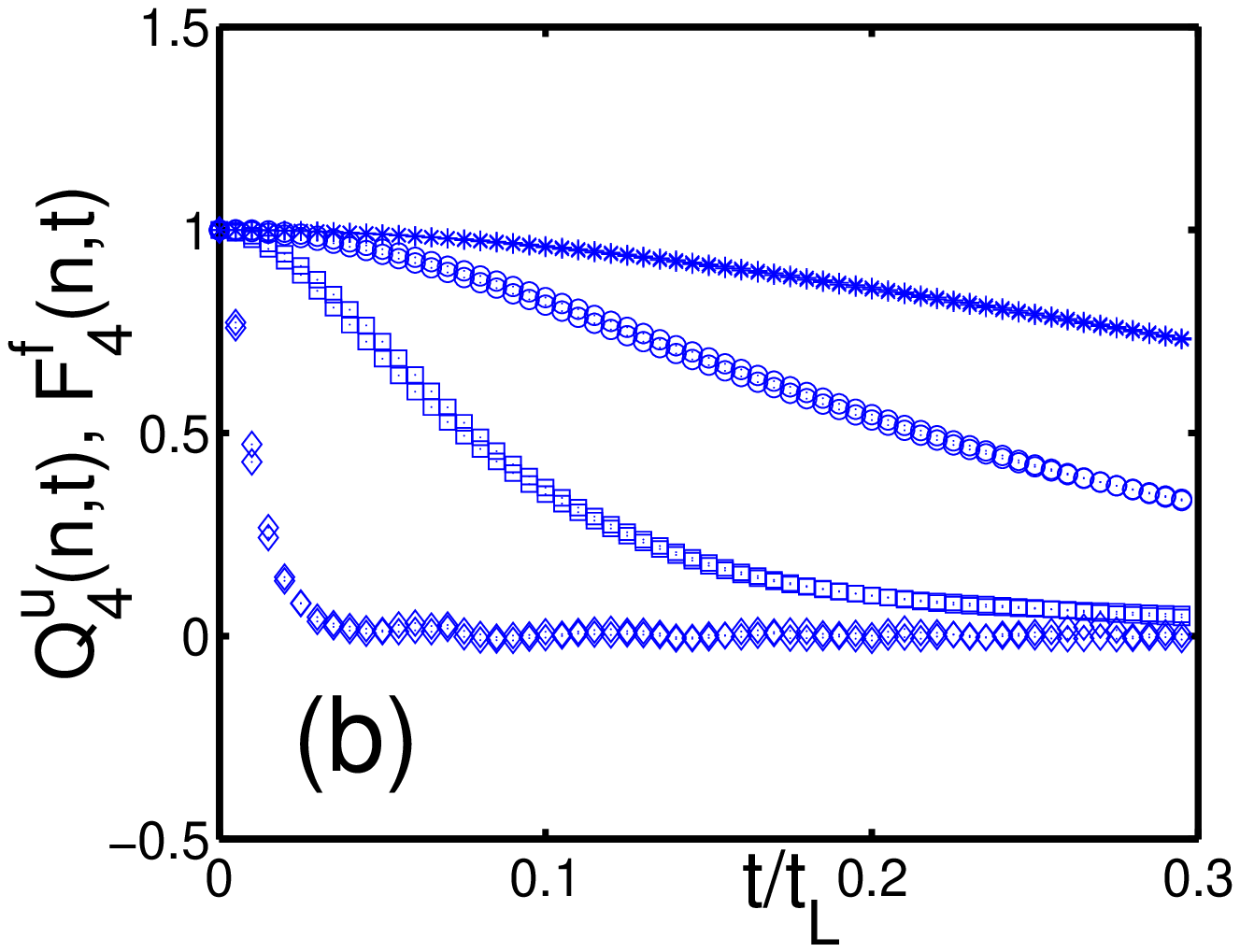} 
\caption{(a) Representative plots of $Q^{u}_p(n,t)$  versus the 
dimensionless time $t/t_L$,  for $p$=2, $n$=9 (lower curve) and 
$p=4$, $n$=5 (upper curve), and 6 different time-origins $t_0$ in 
decaying fluid turbulence for Model C.  Successive time origins are 
separated by 0.5$t_L$. The different symbols for the different sets 
of data for the different time-origins are indistinguishable; this is
compelling numerical evidence for our proposed factorised form of 
the time-dependent structure functions.
\\ 
(b) Plots of $Q^{u}_4(k_n,t)$ and $F^f_4(k_n,t)$ as a function of 
 $t/t_L$ to numerically test our proposed {\it Ansatz} 
(\ref{decoupled2}) decaying turbulence in Model C.   
Results are shown only for shells $n=4$ (uppermost curve), 6, 8, and 
12 (lowermost curve) for clarity.} 
\label{QFcomparison}
\end{center}
\end{figure}

We investigate now whether time-dependent velocity structure 
functions in decaying fluid turbulence show a factorisation similar 
to (\ref{K-QL}) and (\ref{decoup}) for Models A and B. Let us  
normalise the order-$p$, time-dependent structure function by its 
value at $t=0$: 
\begin{equation}
Q_p^{u}(n,t) \equiv \frac{F^u_p(k_n,t_0,t)}{F^u_p(k_n,t_0,0 )};
\label{f2bys2}
\end{equation}
to examine the dependence of $Q_p^{u}(n,t)$ on $t_0$.
 We give, in Figure (\ref{QFcomparison}a), representative plots 
of $Q^{u}_2(n=9,t)$ and $Q^{u}_4(n=5,t)$ versus $t/t_L$ for 6 
different time origins $t_0$; succesive values of $t_0$ are 
separated from each other by $0.5t_L$. To the extent that 
the symbols for different values of $t_0$ overlap, our results 
show that $Q_p^{u}(n,t)$ is independent of $t_0$ (so we have not 
included $t_0$ as an argument of $Q_p^{u}$), i.e., 
$F^u_p(k_n,t_0,t)$ factorises into a part that depends on $t_0$ 
and another that does not [cf., (\ref{K-QL}) and (\ref{decoup})
for the simple, linear Models A and B]. Clearly the 
dynamic-multiscaling exponents extracted from $Q_p^{u}(n,t)$ cannot 
depend on $t_0$. 

Furthermore, these exponents are the same as their counterparts for 
statistically steady, forced turbulence. We illustrate this in
Fig. (\ref{QFcomparison}b) by comparing our numerical 
results for the normalised, time-dependent structure function 
$F^f_4(k_n,t)/S^u_4(k_n)$, for statistically steady, forced
(superscript $f$) turbulence, and $Q^u_4(n,t)$. [To obtain a
statistical steady state for the GOY shell model we use
the external force $f_n = (1 + \imath)\times 5\times 10^{-3}
\delta_{n,1}$.] We have checked explicitly, for $1\leq p \leq 6$, 
that $Q^u_p(n,t)$ and $F_p^f(k_n,t)/S^u_p(k_n)$ agree within our 
error bars. Hence we propose the following factorisation that 
relates the time-dependent structure functions for decaying
and statistically steady (superscript $f$) turbulence,
\begin{equation}
F_p^{u}(k_n,t_0,t) = A_p(k_n,t_0)F_p^f(k_n,t),
\label{decoupled2}
\end{equation}
with all the $t_0$ dependence on the right-hand side in the coefficient 
function $A_p$.
If we now use the multifractal form for $F_p^{f}$ suggested 
in~\cite{mitra1}, we obtain the shell-model analogue of equation (\ref{Fpdecay}). 

\begin{figure}[htbp]
\begin{center}
\includegraphics[height=6cm,width=6cm]{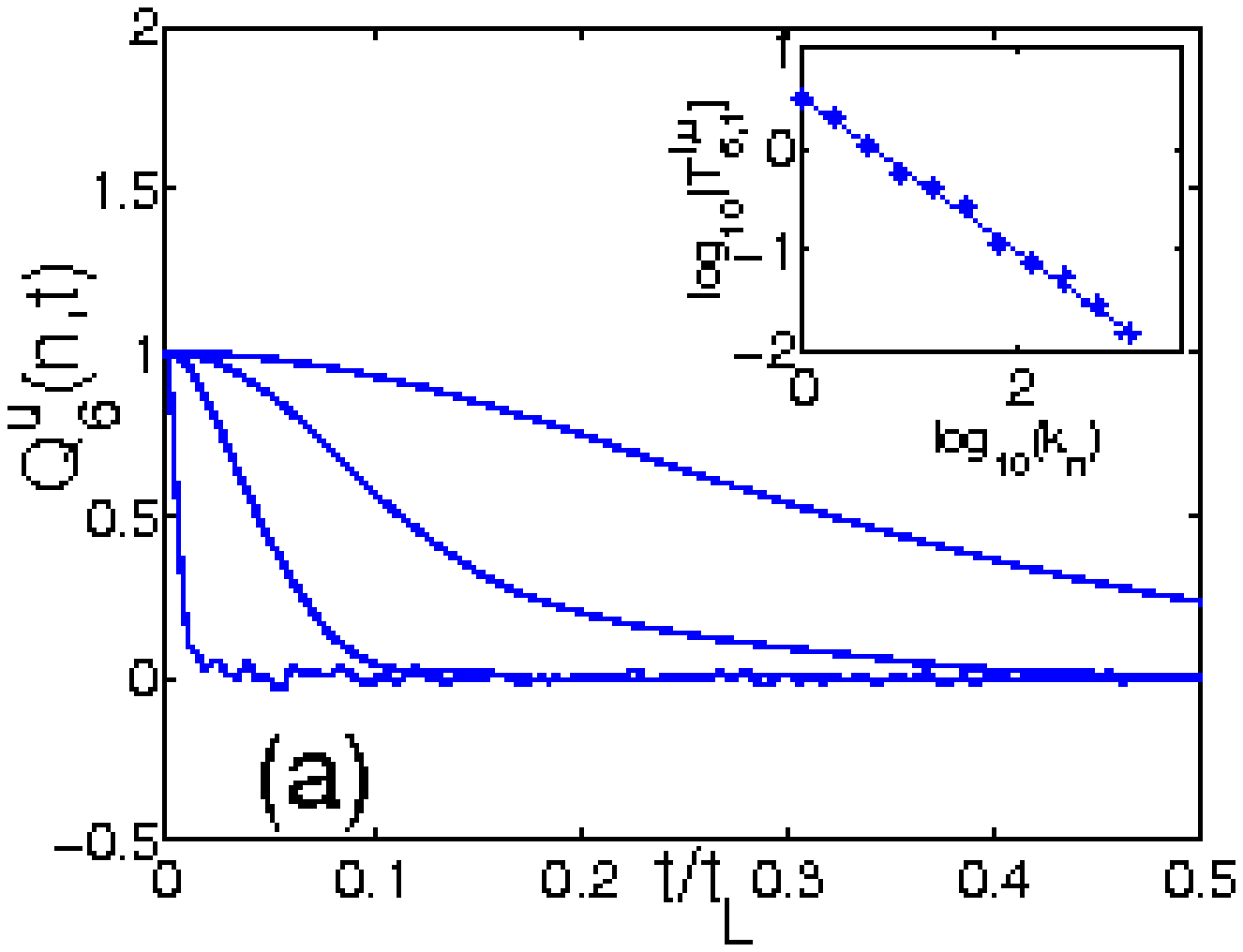}
\includegraphics[height=6cm,width=6cm]{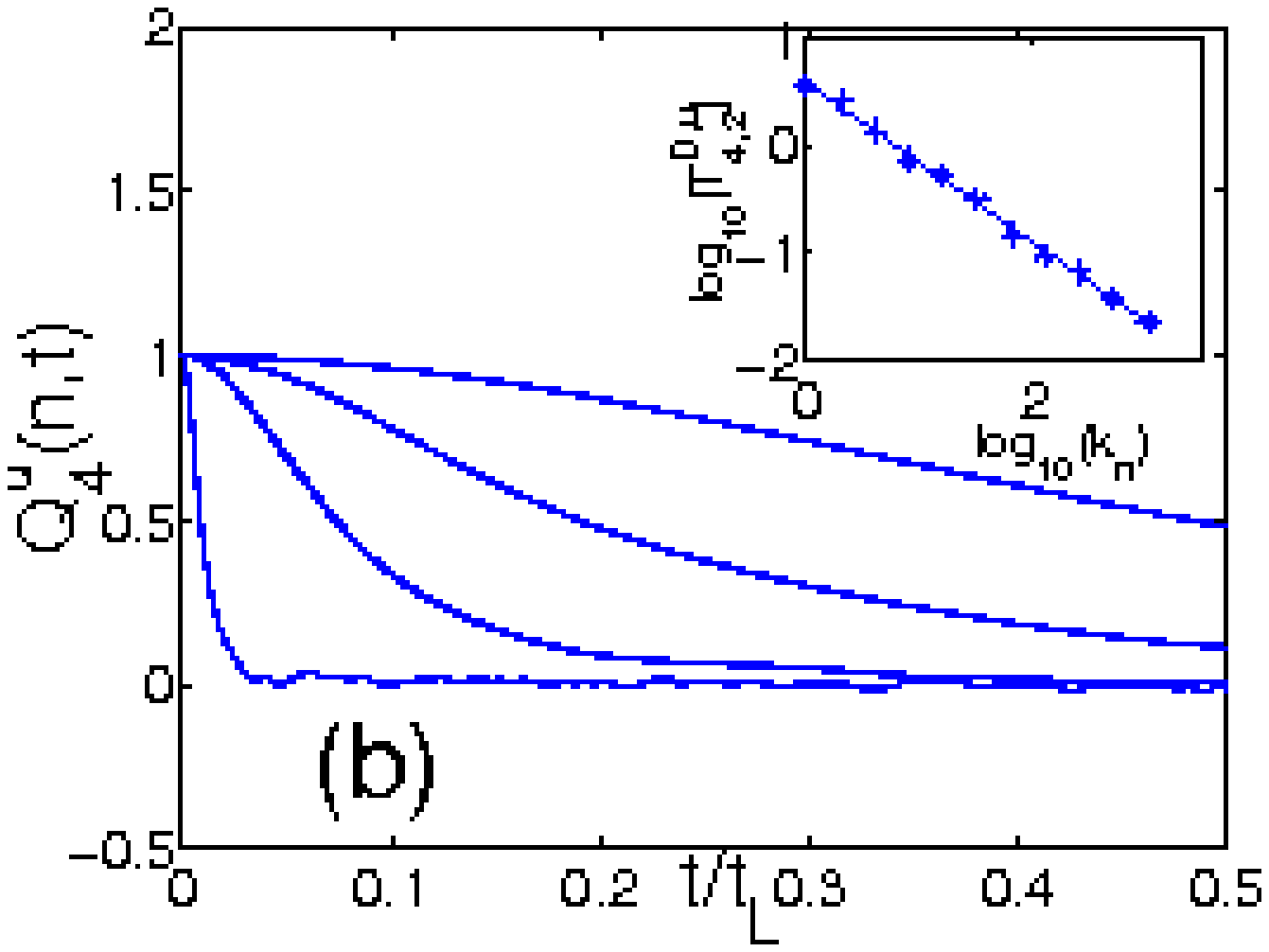} 
\caption{(a) Representative plots of $Q^{u}_6(n,t)$ versus $t/t_L$ 
for Model C; for clarity we show shell numbers 
$n$ = 4 (uppermost), 6, 8 and 12 (lowermost). 
The inset shows $T^{I,u}_{6,1}(n)$ versus $k_n$ on a log-log scale.
A linear fit yields the dynamic-multiscaling exponent 
$z^{I,u}_{6,1} = 0.76 \pm 0.02$. \\
(b) Representative plots of $Q^{u}_4(n,t)$ versus $t/t_L$  
for the same shell numbers as in (a). The inset shows 
$T^{D,u}_{4,2}(n)$ versus $k_n$ on a log-log scale. A linear fit 
yields $z^{D,u}_{4,2} = 0.76 \pm 0.01$.}
\label{Q6}
\end{center}
\end{figure}

We calculate the integral- and the derivative-time scales and their 
associated exponents $z^{I,u}_{p,1}$ and $z^{D,u}_{p,2}$. 
The counterparts of equations (\ref{timp}), for $M = 1$, and (\ref{tdp}), 
 for $M = 2$,  for the GOY model are :
\begin{eqnarray}
T^{I,u}_{p,1}(n) &=& \int_0^{t_\mu}Q^u_p(n,t)dt; \\
\label{goyint}
T^{D,u}_{p,2}(n) &=& \left[\frac{d^2Q^u_p(n,t)}{dt^2}\bigg{|}_{t=0}\right]^{-1/2}; 
\label{goyder}
\end{eqnarray}
here $t_{\mu}$ is the time at which $Q^u_p(n,t) = \mu$, with 
$0\leq \mu \leq 1$. In principle we should use $\mu = 0$, i.e.,
$t_{\mu} = \infty$, but this is not possible in any numerical 
calculation since $Q^u_p$ cannot be obtained accurately for large 
$t$. We use $\mu = 0.6$; and we have checked in representative 
cases that our results do not change for $0.3< \mu <0.7$. To compute
$T^{D,u}_{p,2}(n)$ we use a centred, finite-difference, 
sixth-order scheme. Slopes of log-log plots of $T^{I,u}_{p,1}(n)$ 
and $T^{D,u}_{p,2}(n)$ versus $k_n$ $(4\leq n \leq 14)$ give us 
$z^{I,u}_{p,1}$ (Fig \ref{Q6}a) and $z^{D,u}_{p,2}$ (Fig \ref{Q6}b),
respectively. 

Our results for equal-time and dynamic-multiscaling exponents for 
the GOY model (for both types of initial conditions) are given in 
Tables 1 and 2, respectively. We compute the multiscaling 
exponents for equal-time and time-dependent structure functions for 
50 different cases. Tables 1 and 2 list the means of these values 
and their standard deviations yield the error bars. By comparing
columns 2 of Tables 1 and 2 with column 2 of Table II in 
Ref.~\cite{mitra1} we confirm, for the GOY model, the {\it weak} 
version of universality \cite{lvov1}, i.e., the equal-time  
exponents $\zeta^u_p$ are the same for both decaying and 
statistically steady turbulence. Furthermore, our exponents for the
GOY shell model agree with those presented in Ref.~\cite{lvov1}
for the Sabra shell model.

\begin{figure}[htbp]
\begin{center}
\includegraphics[height=6cm,width=6cm]{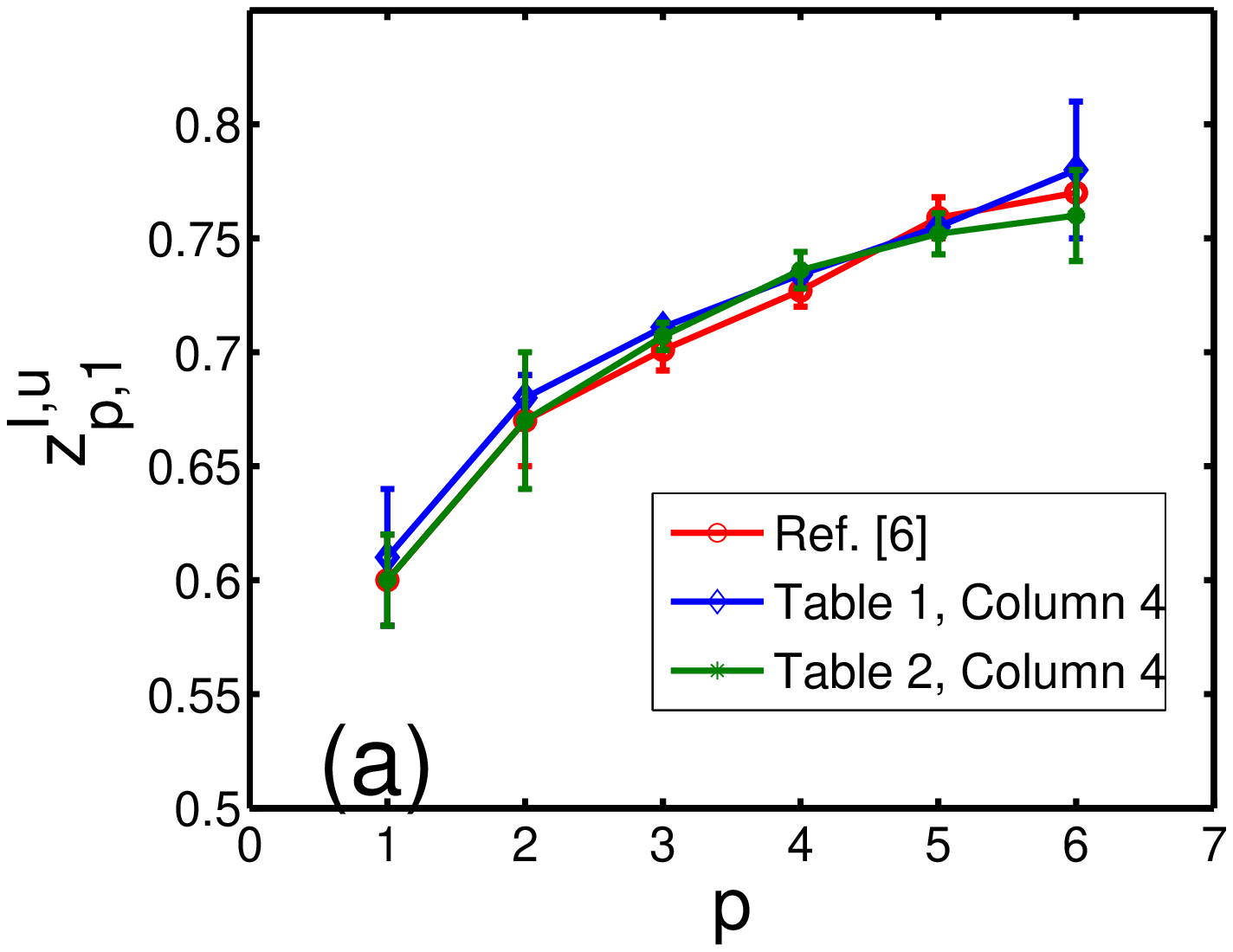}
\includegraphics[height=6cm,width=6cm]{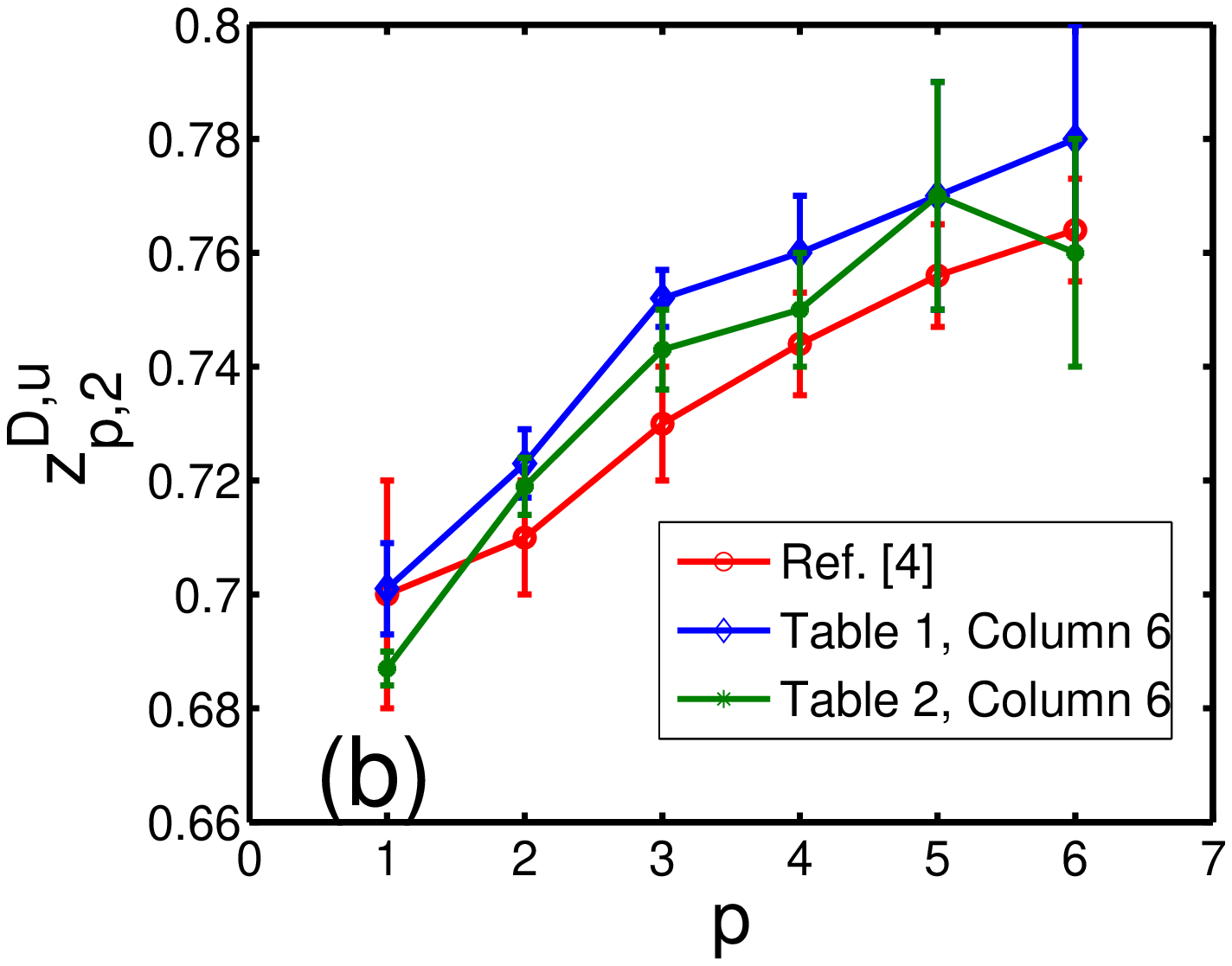} 
\caption{(a) Plots of $z^{I,u}_{p,1}$, with error bars, versus $p$ 
for Model C with data from ~\cite{mitra1}, for statistically steady 
turbulence, and columns 4 of Tables 1 and 2, for decaying 
turbulence; these plots illustrate the agreement between the three 
sets of exponents. \\ 
(b) We compare the derivative-time exponents $z^{D,u}_{p,2}$ from 
~\cite{mitra1} and columns 6 of Tables 1 and 2. 
As in (a), we find that the three sets of exponents agree, within 
error bars, with each other.}
\label{comparison1}
\end{center}
\end{figure}

\begin{figure}[htbp]
\begin{center}
\includegraphics[height=6cm,width=6cm]{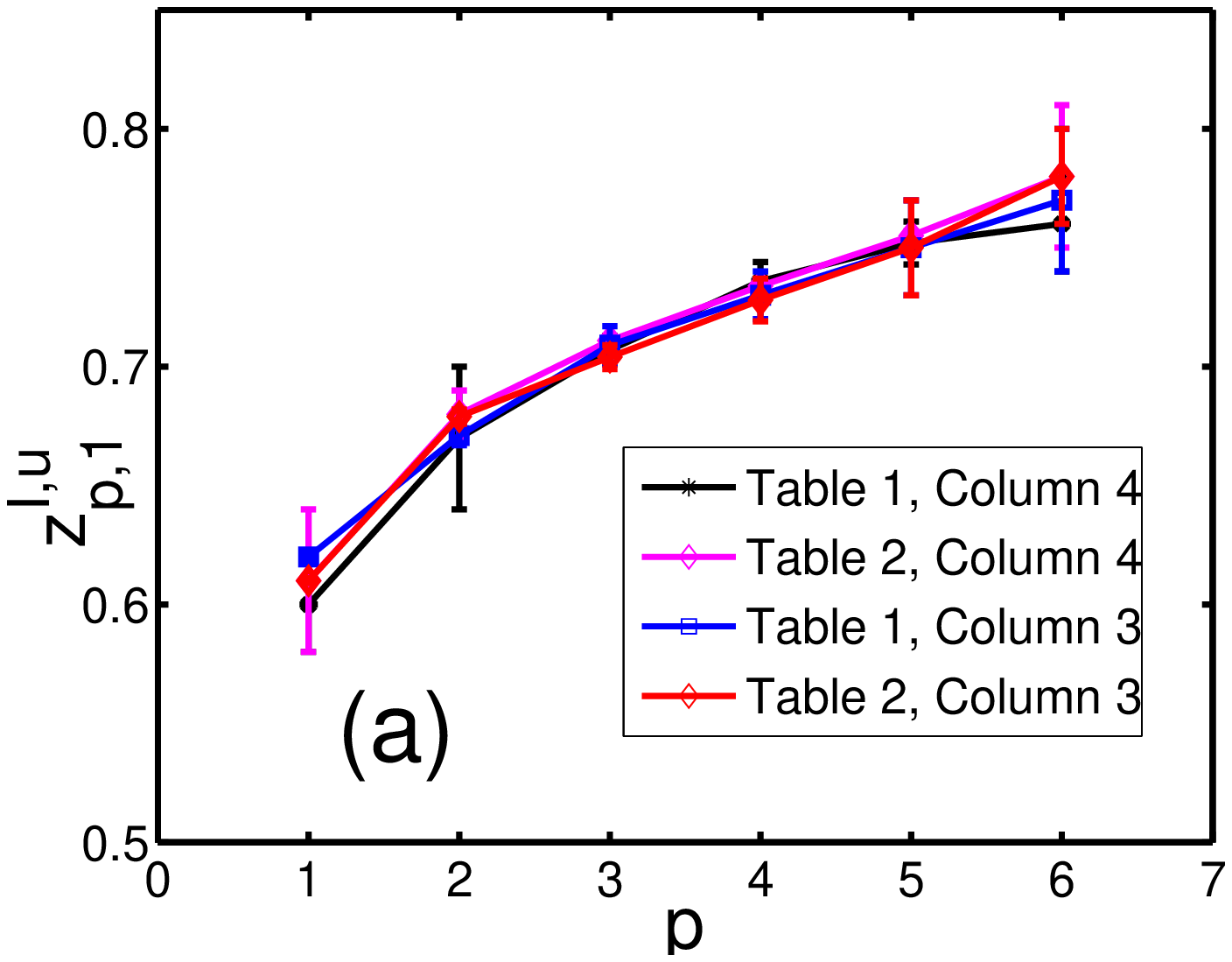}
\includegraphics[height=6cm,width=6cm]{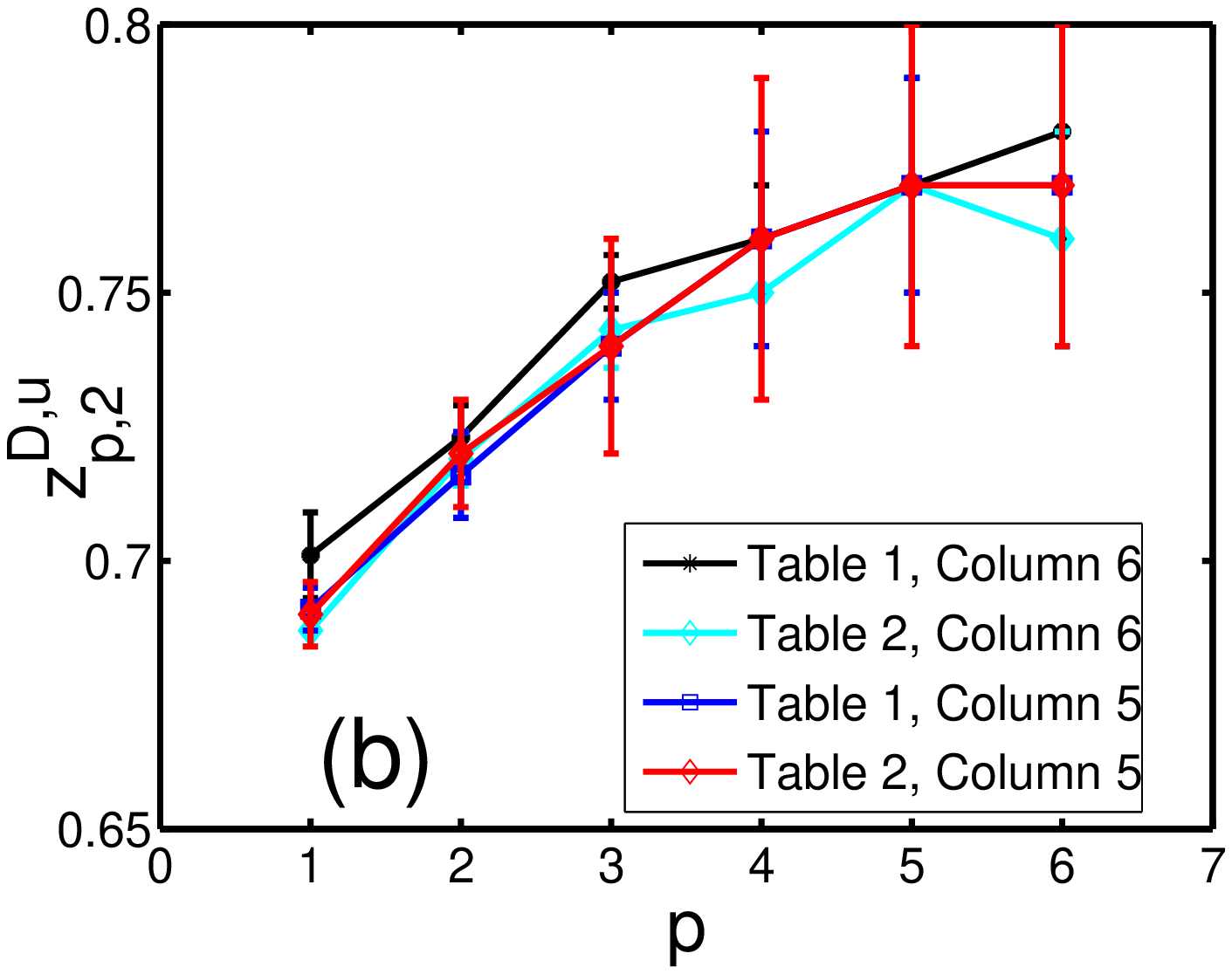} 
\caption{(a) Plots of $z^{I,u}_{p,1}$, with error bars, versus $p$ 
for Model C with values obtained via the bridge relations (columns 3, Tables 1 and 2) 
and those obtained from our numerical simulations (columns 4, Tables 1 and 2). \\ 
(b)Plots of $z^{D,u}_{p,2}$, with error bars, versus $p$ 
for Model C with values obtained via the bridge relations (columns 5, Tables 1 and 2) 
and those obtained from our numerical simulations (columns 6, Tables 1 and 2).} 
\label{comparison2}
\end{center}
\end{figure}

A comparison of the remaining columns of Tables 1 and 2 with their 
counterparts in Table II of Ref.~\cite{mitra1} shows that this weak 
universality also applies to dynamic multiscaling exponents.
Moreover, our direct numerical results for $z^{I,u}_{p,1}$ and 
$z^{D,u}_{p,2}$ (columns $4$ and $6$ in Tables 1 and 2) agree with 
the bridge-relation values of these exponents (columns $3$ and $5$ 
in Tables 1 and 2) that follow from equations (\ref{zipm}) and 
(\ref{zdpm}) and $\zeta^{u}_p$ (columns 2 of Tables 1 and 2). Note 
that the agreement between corresponding entries in Tables 1 and 2 
shows that our results are insensitive to the type of initial 
conditions we use. Finally, if we compare these Tables with Table 
II in Ref. ~\cite{mitra1}, we find that our dynamic-multiscaling 
exponents for {\it decaying}, homogeneous, isotropic turbulence 
agree with their counterparts for the {\it statistically steady}
case. Pictorial comparisons of the data in Tables 1 and 2 and the 
results of Ref.~\cite{mitra1} are shown in Figs. 
(\ref{comparison1}a) and (\ref{comparison1}b) for integral-time 
(columns $4$ in Tables 1 and 2) and derivative-time 
(columns $6$ in Tables 1 and 2) exponents.
Similarly Figs. (\ref{comparison2}a) and (\ref{comparison2}b)
compare the dynamic-multiscaling exponents from our direct
numerical simulations with the values predicted for them by
the bridge relations (\ref{zipm}) and (\ref{zdpm}) and the 
equal-time exponents $\zeta_p^u$ given in Tables 1 and 2.

\begin{table*}
\framebox{\begin{tabular}{c|c|c|c|c|c}
order$(p)$ & $\zeta^{u}_p$ & $z^{I,u}_{p,1}$[Eq.(\ref{zipm})] & $z^{I,u}_{p,1}$
 & $z^{D,u}_{p,2}[Eq.(\ref{zdpm})]$ & $z^{D,u}_{p,2}$  \\
\hline
 1 &  0.379 $\pm$ 0.008 & 0.621 $\pm$ 0.008  & 0.61 $\pm$ 0.03
    & 0.68  $\pm$  0.01 & 0.699 $\pm$ 0.008
   \\
 2 &  0.711 $\pm$ 0.002  & 0.66 $\pm$ 0.01 & 0.68 $\pm$ 0.01
   & 0.716 $\pm$ 0.008  & 0.723  $\pm$  0.006
 \\
 3 &  1.007 $\pm$ 0.003  & 0.704 $\pm$ 0.005 & 0.711 $\pm$ 0.001
   & 0.74 $\pm$ 0.01  & 0.752 $\pm$ 0.005
  \\
 4 &  1.279 $\pm$ 0.006  &  0.728 $\pm$ 0.009 & 0.734 $\pm$ 0.002
 & 0.76 $\pm$ 0.02  & 0.76  $\pm$ 0.01  \\

 5 &  1.525 $\pm$ 0.009 & 0.75 $\pm$ 0.02  & 0.755 $\pm$ 0.002
   & 0.77  $\pm$ 0.02  & 0.77 $\pm$ 0.02
  \\
 6 &  1.74  $\pm$ 0.01 & 0.78  $\pm$ 0.02  & 0.78 $\pm$ 0.03
   & 0.77  $\pm$ 0.03   & 0.78 $\pm$ 0.02
   \\

\end{tabular}}

\caption{Our simulation results for Model C with Type I initial 
conditions. Order-$p$ (column 1); equal-time 
exponents $\zeta^{u}_p$
(column 2); integral-scale dynamic-multiscaling exponent
$z^{I,u}_{p,1}$ (column 3) from the bridge relation (\ref{zipm})
and the values of $\zeta^{u}_p$ in column 2; $z^{I,u}_{p,1}$ from our
calculation using time-dependent structure functions (column 4);
the derivative-time exponents
$z^{D,u}_{p,2}$ (column 6) from the bridge relation (\ref{zdpm}) and the values of
$\zeta^{u}_p$ in column 2; $z^{D,u}_{p,2}$ from our calculation
using time-dependent structure function (column 7).
The error estimates are obtained as described in the text.}
\end{table*}

\vspace{0.5cm}

\begin{table*}
\framebox{\begin{tabular}{c|c|c|c|c|c}
order$(p)$ & $\zeta^{u}_p$ & $z^{I,u}_{p,1}$[Eq.(\ref{zipm})] & $z^{I,u}_{p,1}$
 & $z^{D,u}_{p,2}[Eq.(\ref{zdpm})]$ & $z^{D,u}_{p,2}$  \\
\hline
 1 &  0.380 $\pm$ 0.004  & 0.620 $\pm$ 0.004 & 0.60 $\pm$ 0.02
    &0.690  $\pm$ 0.009 & 0.687 $\pm$ 0.003
   \\
 2 &  0.709 $\pm$ 0.003  & 0.671 $\pm$ 0.007 & 0.67 $\pm$ 0.03
   & 0.72 $\pm$ 0.01  & 0.719 $\pm$ 0.005
 \\
 3 &  1.000 $\pm$ 0.005  & 0.709 $\pm$ 0.008 & 0.707 $\pm$ 0.006
   & 0.74 $\pm$ 0.02  & 0.743 $\pm$ 0.007
  \\
 4 &  1.266 $\pm$ 0.008  & 0.73 $\pm$ 0.01 & 0.736 $\pm$ 0.008
 & 0.76 $\pm$ 0.03  & 0.75 $\pm$ 0.01   \\

 5 &  1.51 $\pm$ 0.01  & 0.75 $\pm$ 0.02  & 0.752 $\pm$ 0.009
   &0.77  $\pm$ 0.03  & 0.77 $\pm$ 0.02
  \\
 6 &  1.74 $\pm$ 0.02   & 0.77 $\pm$ 0.03   & 0.76 $\pm$ 0.02
   & 0.77  $\pm$ 0.03   & 0.76 $\pm$ 0.02
   \\
\end{tabular}}
\caption{Our simulation results for Model C with Type II initial conditions. 
Order-$p$ (column 1); equal-time exponents $\zeta^{u}_p$
(column 2); integral-scale dynamic-multiscaling exponent
$z^{I,u}_{p,1}$ (column 3) from the bridge relation (\ref{zipm})
and the values of $\zeta^{u}_p$ in column 2; $z^{I,u}_{p,1}$ from our
calculation using time-dependent structure functions (column 4);
the derivative-time exponents
$z^{D,u}_{p,2}$ (column 6) from the bridge relation (\ref{zdpm}) and the values of
$\zeta^{u}_p$ in column 2; $z^{D,u}_{p,2}$ from our calculation
using time-dependent structure function (column 7).
The error estimates are obtained as described in the text.}
\label{data-table}
\end{table*}

\subsection{Model D} 
\label{d}
\begin{figure}[htbp]
\begin{center}
\includegraphics[height=6cm,width=6cm]{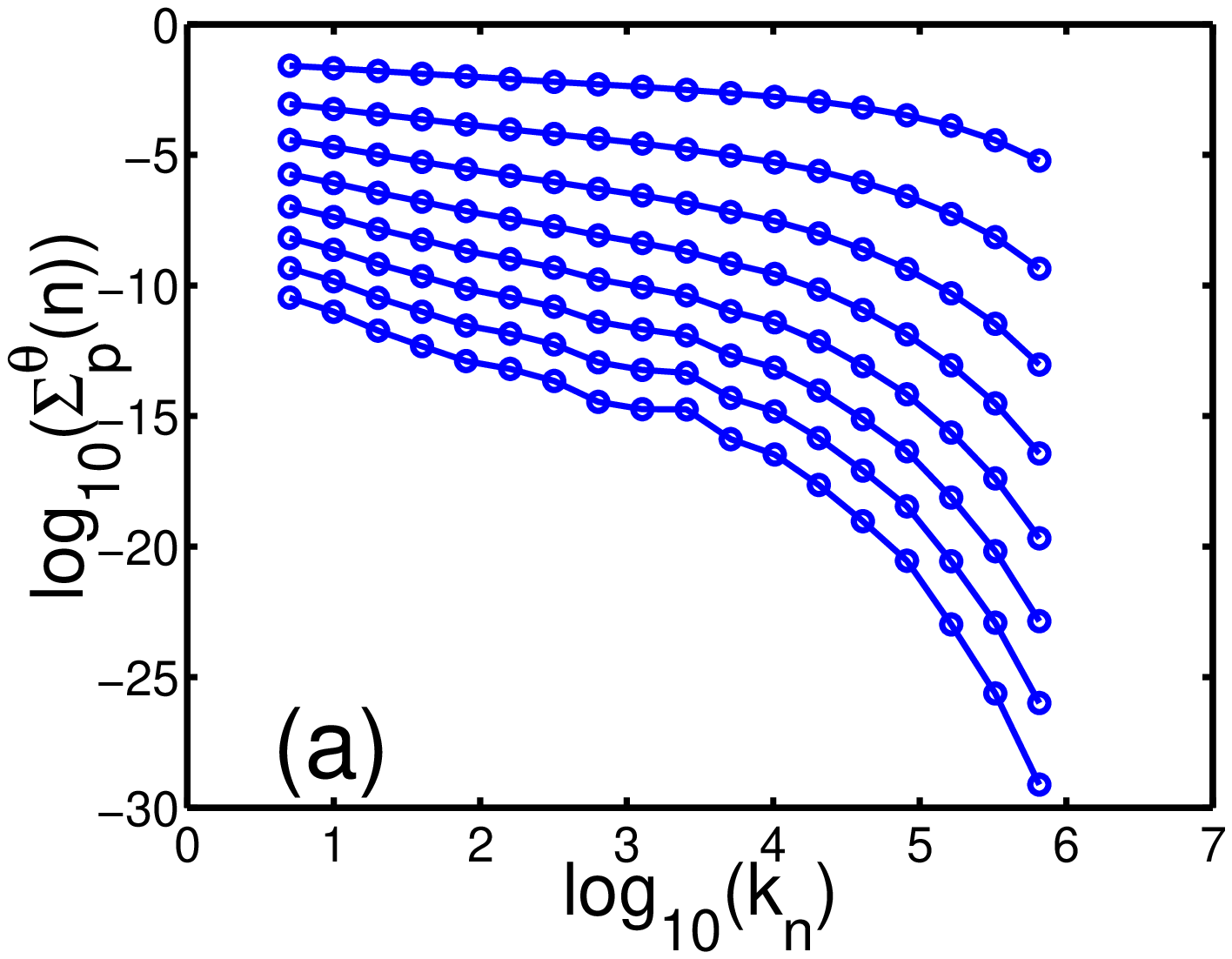}
\includegraphics[height=6cm,width=6cm]{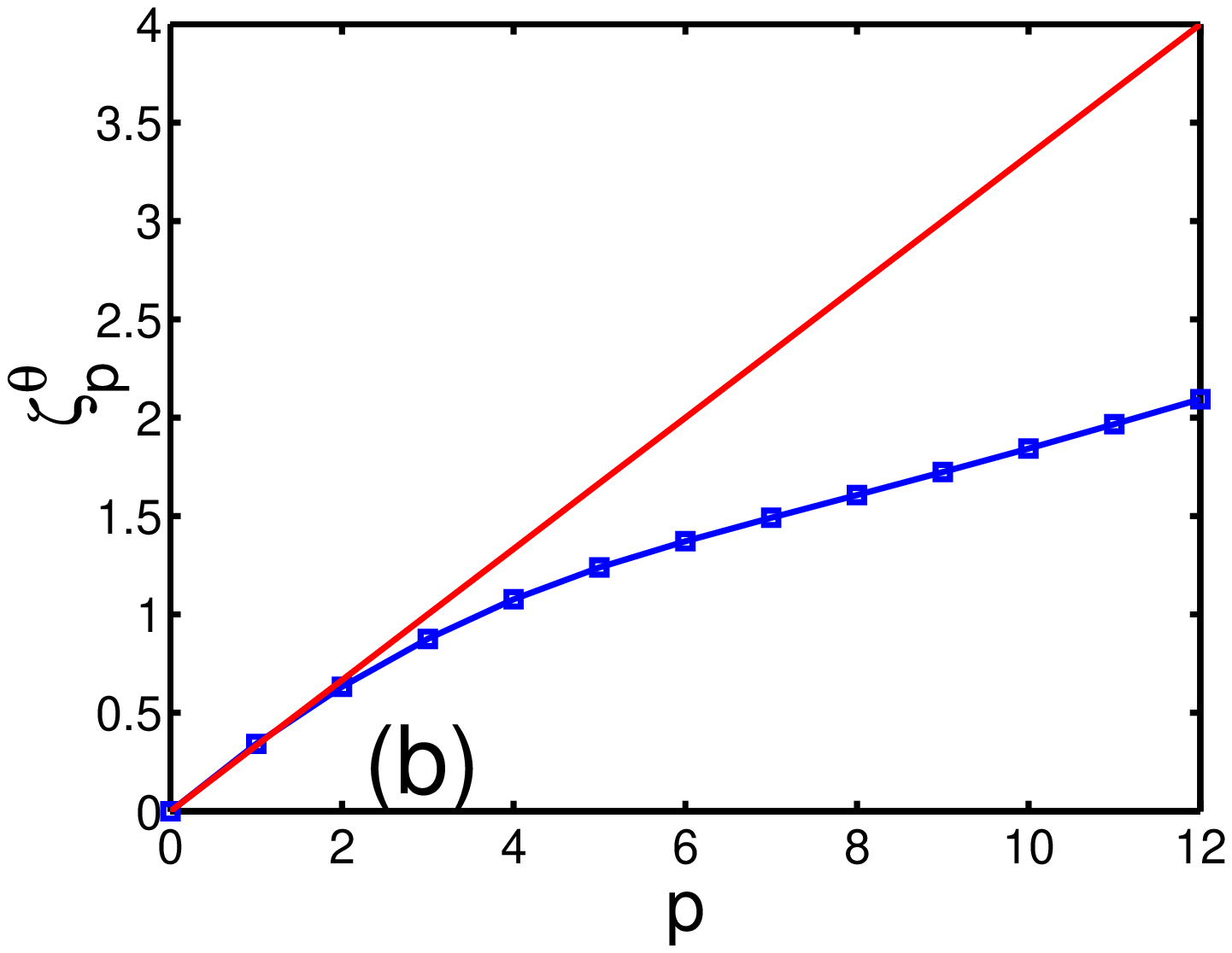}  
\caption{(a) Representative plots of $\Sigma_p^\theta(n)$ versus 
$k_n$ on a logarithmic scale for $p$ = 1 (uppermost curve) to 8 
(lowermost curve) for Model D.
\\(b) Plot of $\zeta^{\theta}_p$, obtained from the linear, inertial 
region in (a), versus $p$. Our data points, shown as squares,
are connected by a line; the error bars are smaller than the size of 
the symbol. The straight line corresponds to the Kolmogorov 
prediction  of $p/3$.}  
\label{statD}
\end{center}
\end{figure}

From a numerical study of a passive-scalar shell model, with
advecting velocities from the Sabra shell model, it was 
shown~\cite{arad,cohen} that the equal-time scaling exponents 
$\zeta_p^{\theta}$ are universal: they are the same for decaying 
and statistically steady turbulence; and, in the latter case, they 
do not depend on the type of forcing. We find, not surprisingly, 
that this universality holds even when the advecting velocity field 
is a solution of the GOY model, i.e., for Model D:  Table 3 column 2
shows the values of $\zeta_p^{\theta}$ for $1 \leq p \leq 6$,
which we have obtained for decaying turbulence; these agree with 
the results of Refs.~\cite{mitra2,jensen} for statistically steady 
turbulence in Model D. Our equal-time exponents are also within 
error bars of their counterparts for the passive-scalar shell model
of Refs.~\cite{arad,cohen}. We obtain $\zeta_p^{\theta}$ from
log-log plots such as Fig. (\ref{statD}a) for the modified, equal-time 
structure function (\ref{sigmap}) $\Sigma_p^\theta$ versus $k_n$; 
the slope of the linear region $4 \leq n\leq 12$ yields 
$\zeta_p^{\theta}$ that is plotted versus $p$ in Fig. (\ref{statD}b). 

\begin{figure}[htbp]
\begin{center}
\includegraphics[height=6cm,width=6cm]{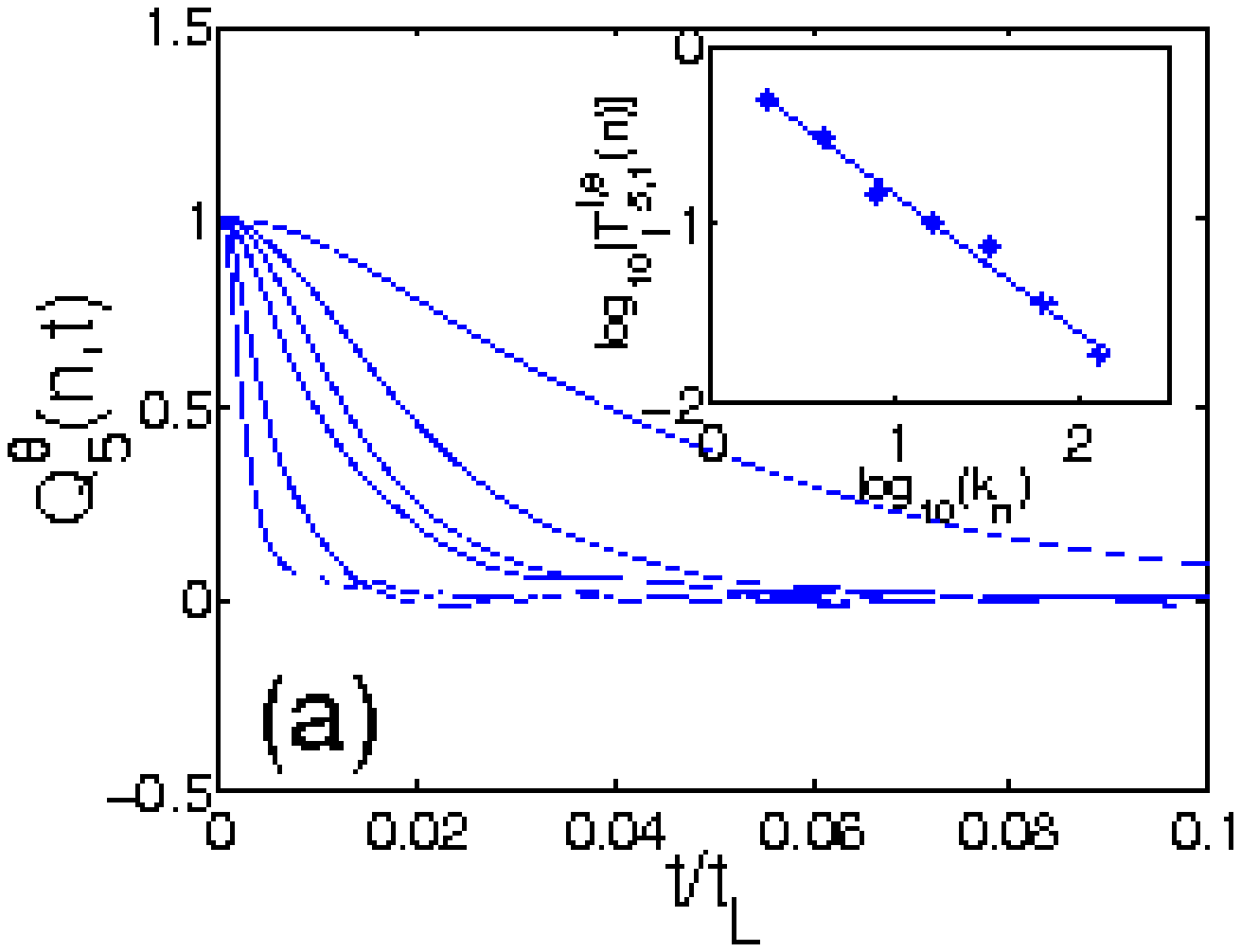} 
\includegraphics[height=6cm,width=6cm]{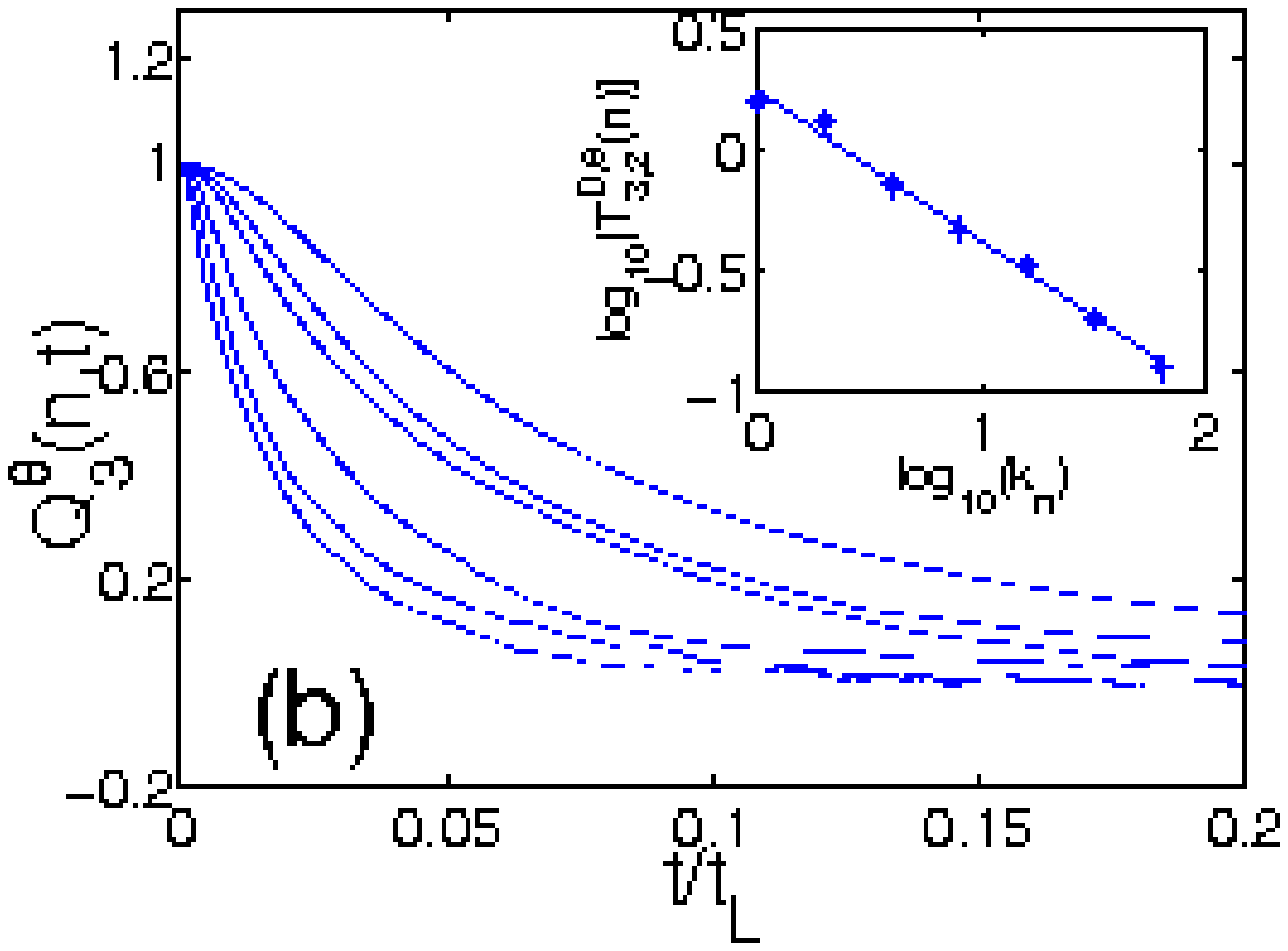} 
\caption{(a) Representative plots of $Q^{\theta}_5(n,t)$ 
versus $t/t_L$, for shell numbers 6 (uppermost) to 11 (lowermost) 
for Model D. The inset shows a log-log plot of 
$T^{I,\theta}_{5,2}(n)$ versus $k_n$. A linear fit 
yields $z^{I,\theta}_{5,1} = 0.562 \pm 0.006$.
  \\
(b) Representative plots of $Q^{\theta}_3(n,t)$ versus $t/t_L$ 
for the same shell numbers as in (a). 
The inset shows $T^{D,\theta}_{3,2}(n)$ versus $k_n$ on a log-log 
scale. A linear fit yields $z^{D,\theta}_{3,2} = 0.646 \pm 0.003$.}
\label{integ5}
\end{center}
\end{figure}

To analyse the time-dependent, passive-scalar structure 
functions we follow $\Sref{c}$ for Model C: We obtain 
integral- and derivative-time scales from equations 
(\ref{timp}) and (\ref{tdp}), for $M$ = 1 and $M$ = $2$, 
respectively. In the integral in (\ref{timp}) we set the upper 
limit to $t_\mu$, the time at which the normalised, time-dependent 
structure function 
\begin{equation}
Q_p^{\theta}(n,t)\equiv \frac{F^{\theta}_p(k_n,t_0,t)}{F^{\theta}_p
(k_n,t_0,0)} 
\end{equation}
is equal to $\mu$, with $0 \leq \mu \leq 1$. We use $\mu = 0.6$, 
but we have checked in representative cases that our results remain 
unchanged if we use the range of values $0.3 \leq \mu \leq 0.8$. 
Slopes of log-log plots of $T^{I,\theta}_{p,1}(n)$ versus $k_n$ 
yield $z^{I,\theta}_{p,1}$. To extract the derivative-time scale, 
we use a centred, finite-difference, sixth-order scheme to obtain 
$T^{D,\theta}_{p,2}(n)$ from which we get $z_{p,2}^{D,\theta}$.  
Integral- and derivative-time multiscaling exponents are extracted 
from linear fits in the inertial range $4 \leq n \leq 10$ as 
shown in the insets of the representative plots in  
Figs. (\ref{integ5}a) and (\ref{integ5}b) for, respectively,
$Q_5^{\theta}(n,t)$ and $Q_3^{\theta}(n,t)$ versus $t/t_L$. 
Finally, from the bridge relations (\ref{eq:pas:zp_pshell}) 
and our GOY-model, equal-time exponents 
$\zeta_{-1}^u = -0.44 \pm 0.04$ and $\zeta_2^u = 0.709 \pm 
0.003$ we obtain $z^{I,\theta}_{p,1} = 0.56 \pm 0.04$ and 
$z^{D,\theta}_{p,2} = 0.645 \pm 0.003$ in agreement with
the values from our simulations listed in columns 3 and 4 of
Table 3. By comparing these columns with their counterparts in 
Table II of Ref. ~\cite{mitra2}, we find agreement, within our 
error bars, between the dynamic-multiscaling exponents for Model
D for both statistically steady and decaying turbulence.

\section{Conclusions}
\label{conclusion}
We have systematised the study of the dynamic multiscaling 
of time-dependent structure functions in four Models (A-D) for
passive-scalar (A,B, and D) and fluid (C) turbulence.
By a suitable normalisation of these structure functions,
we eliminate their dependence on the origin of time $t_0$ at which
we initiate our measurements. We have shown analytically that for the 
Kraichnan model of passive-scalar turbulence (Models A and B) 
the two-point time-dependent structure function can be factorised  
into two parts, one depending on the origin of time $t_0$ and another 
that does not. This suggests a suitable normalisation of the 
structure functions by which we can eliminate
their dependence on $t_0$. Surprisingly the same normalisation works for other models
of fluid turbulence (Model C) and passive-scalar turbulence (Model D) as we have 
shown by extensive numerical simulations. Once this 
dependence on $t_0$ has been factored out, the methods developed earlier~\cite{mitra1,mitra2}
for statistically steady turbulence yield linear bridge relations
that connect dynamic-multiscaling and equal-time exponents.
We show analytically, for Models A and B, and numerically, 
for Models B, C, and D, that these exponents and bridge relations
are the same for statistically steady and decaying turbulence.      
Thus we have generalised the universality of equal-time 
exponents~\cite{lvov1} and provided strong evidence for dynamic 
universality, i.e., dynamic-multiscaling exponents do not depend on 
whether the turbulence decays or is statistically steady for each 
of the Models A-D. We have shown elsewhere that these exponents
do not depend on the dissipation mechanism by studying the 
GOY model with a hyperviscous term ~\cite{sp23}.

It is useful to distinguish our results from other studies
of temporal dependences of quantities in turbulence. Two such
results are described below.

The first example is the result $E_{tot}(t) \sim t^{-2}$, which 
holds for the total energy $E_{tot}$ at large times $t$ in decaying 
turbulence ~\cite{frisch}. This result holds for times that are longer 
than the time at which the integral length scale becomes comparable to 
the size of the system. [In decaying turbulence, the integral scale
$L_{int}(t) \equiv (\int dk E(k,t)/k)/(\int dk E(k,t))$ increases
with time since the large-$k$ part of the energy spectrum $E(k,t)$,
at time $t$, gets depleted as $t$ increases.] In all the results
we report here, our shell-model analogue of $L_{int}$ is well below
the linear size of the system, i.e., $L_{int} \ll k_0^{-1}$.
Thus our results are not modified significantly by finite-size
corrections, nor does the trivial decay of $E_{tot}$, mentioned
above, set in and  mask the dynamic multiscaling we have elucidated. 

The second example is the intermittency of velocity time increments
studied in Ref.~\cite{chevi}. These studies calculate structure functions 
along a single Lagrangian trajectory. Spatial separations, of the 
sort used in defining equal-time Eulerian structure functions,
are replaced by temporal separations along a Lagrangian trajectory.
This is distinct from the spatiotemporal structure functions we
consider and which are required for a full elucidation
of dynamic multiscaling here (and analogous dynamic scaling 
in critical phenomena).

We have described the way in which we have obtained error bars 
for the equal-time and dynamic-multiscaling exponents. These 
error bars only account for statistical errors but not the 
systematic errors associated with the values of $n$ over which
we fit inertial-range exponents. One can try to estimate such 
systematic errors by obtaining local slopes of log-log plots
that yield these exponents. However, local slopes can be deceptive
in shell models since the values of $k_n$ are separated by factors
of 2. Instead, we can try to estimate these systematic errors
by comparing the exponents we obtain by fitting over the 
ranges $3\leq n \leq 15$, $4 \leq n \leq 14$, and $5 \leq n \leq 13$. 
We have carried out such checks in representative cases for the 
dynamic-multiscaling exponents we report. The error bars we then obtain are
about a factor of 4 larger than those shown in Tables 1-3.

We hope our work will stimulate experimental studies of dynamic
multiscaling in turbulence. Recent advances in the experimental 
techniques of particle tracking in turbulent 
flows~\cite{boden,pinton} have made it possible to obtain 
accurate measurements of Lagrangian properties.  To obtain the 
types of structure functions that we have described it will be
necessary at least to track two Lagrangian trajectories of particles
that are separated initially by a distance $r$. At the level of
second-order structure functions this has been attempted in 
the direct numerical simulations of Ref.~\cite{kaneda}.

\ack{ We would like to thank P. Perlekar for discussions, DST and UGC (India)
for support, and SERC (IISc) for computational resources. One of us 
(RP) is a member of International Collaboration for Turbulence Research.}

\begin{table*}
\begin{center}
\framebox{\begin{tabular}{c|c|c|c}
order$(p)$ & $\zeta_p^{\theta}$ & $z^{I,\theta}_{p,1}$ & $z_{p,2}^{D,\theta}$  \\
\hline
1 &  0.342 $\pm$ 0.002   &  0.522 $\pm$ 0.002 &  0.632 $\pm$ 0.003
 \\
 2 &  0.634 $\pm$ 0.003  &  0.531 $\pm$ 0.004 &  0.647 $\pm$ 0.003
 \\
 3 &   0.873 $\pm$ 0.003  & 0.553 $\pm$ 0.006 & 0.646 $\pm$ 0.003
  \\
 4 &  1.072 $\pm$ 0.004  & 0.563 $\pm$ 0.003 & 0.642 $\pm$ 0.005
   \\
 5 &  1.245 $\pm$ 0.004  & 0.562 $\pm$ 0.006 & 0.643 $\pm$ 0.006
  \\
 6 &  1.370 $\pm$ 0.006  & 0.576 $\pm$ 0.006 & 0.640 $\pm$ 0.005
   \\
\end{tabular}}
\caption{Our simulation results for Model D.
Order-$p$ (column 1); equal-time exponents $\zeta^{\theta}_p$
(column 2); integral-scale dynamic-multiscaling exponent
$z^{I,\theta}_{p,1}$ (column 3); the derivative-time exponents
$z^{D,\theta}_{p,2}$ (column 4). The error estimates are obtained as 
described in the text.}
\label{data-table2}
\end{center}
\end{table*}


\begin{thebibliography}{11}
\bibitem{foot1} In our studies of decaying turbulence, 
we restrict ourselves to initial conditions that lead to a 
cascade of energy from large to small length scales, where
viscous dissipation becomes significant. Certain types of 
power-law initial conditions\cite{chirag} do not lead to 
such a cascade and lead to trivial scaling.
\end{thebibliography}
\end{document}